\RequirePackage{fix-cm}
\documentclass[smallextended]{svjour3}       
\smartqed  
\usepackage{natbib}
\usepackage{balance}
\usepackage{graphicx}
\usepackage{booktabs}
\usepackage{tikz}
\usepackage{multirow}
\usepackage{subfig}
\usepackage{tablefootnote}

\begin{document}

\title{An Exploratory Study on the Introduction and Removal of Different Types of Technical Debt}

\titlerunning{Introduction and Removal of Different Types of Technical Debt}        

\author{Jiakun Liu 
        \and Qiao Huang
        \and Xin Xia 
        \and Emad Shihab 
        \and David Lo   
        \and Shanping Li
        }

\institute{Jiakun Liu \at
                College of Computer Science and Technology, Zhejiang University, China\\
                PengCheng Laboratory, China\\
         \email{jkliu@zju.edu.cn}
         \and
         Qiao Huang \at
                College of Computer Science and Technology, Zhejiang University, China\\
         \email{tkdsheep@zju.edu.cn}
         \and
         Xin Xia \at
                Faculty of Information Technology, Monash University, Melbourne, Australia\\
         \email{xin.xia@monash.edu}
           \and
        Emad Shihab \at
         Department of Computer Science and Software Engineering, Concordia University,  Canada\\
          \email{eshihab@encs.concordia.ca}
         \and
         David Lo \at
                School of Information Systems, Singapore Management University, Singapore\\
         \email{davidlo@smu.edu.sg}
         \and
         Shanping Li \at
         College of Computer Science and Technology, Zhejiang University, China\\
         \email{shan@zju.edu.cn}
        }

\date{Received: date / Accepted: date}

\maketitle

\begin{abstract}
To complete tasks faster, developers often have to sacrifice the quality of the software.
Such compromised practice results in the increasing burden to developers in future development.
The metaphor, technical debt, describes such practice.
Prior research has illustrated the negative impact of technical debt, and many researchers investigated how developers deal with a certain type of technical debt.
However, few studies focused on the removal of different types of technical debt in practice. 

To fill this gap, we use the introduction and removal of different types of self-admitted technical debt (i.e., SATD) in 7 deep learning frameworks as an example.
This is because deep learning frameworks are some of the most important software systems today due to their prevalent use in life-impacting deep learning applications.
Moreover, the field of the development of different deep learning frameworks is the same, which enables us to find common behaviors on the removal of different types of technical debt across projects.
By mining the file history of these frameworks, we find that design debt is introduced the most along the development process.
As for the removal of technical debt, we find that requirement debt is removed the most, and design debt is removed the fastest.
Most of test debt, design debt, and requirement debt are removed by the developers who introduced them.
Based on the introduction and removal of different types of technical debt, we discuss the evolution of the frequencies of different types of technical debt to depict the unresolved sub-optimal trade-offs or decisions that are confronted by developers along the development process.
We also discuss the removal patterns of different types of technical debt, highlight future research directions, and provide recommendations for practitioners.
\end{abstract}

\maketitle
\section{Introduction}
During the process of software development, developers are expected to continuously deliver high-quality products or services.
However, because of time pressure, market competition, and cost reduction \citep{lim2012balancing}, developers are often confronted with a dilemma: a shorter completion time or better software quality.
The compromised decision leads to the increasing burden in the future development life cycle.
The metaphor, technical debt, first proposed by \citet{cunningham1993wycash}, describes such the decision.

Previous research finds that technical debt is detrimental, e.g., increasing the cost, and negatively impacting the product quality \citep{wehaibi2016examining, design_debt_impact, fontana2012investigating}.
Therefore, many researchers investigate the types of technical debt \citep{alves2014towards, identify_td_effectively, li2015systematic} and inspect how developers deal with technical debt \citep{design_debt_impact, ernst2015measure, klinger2011enterprise, spinola2013investigating}.
However, previous research only focused on certain types of technical debt, e.g., observing the introduction and removal of code smell to track how developers deal with design debt \citep{design_debt_impact}.
One of the reasons is that observing technical debt is difficult and requires a thorough analysis of the whole project as the technical debt is often not directly visible.
\textbf{None of them characterize the removal of different types of technical debt at the same time.}

To fill this gap, we use the self-admitted technical debt (i.e., SATD) as an indicator of technical debt.
This is because previous research finds that most of the developers do not consider technical debt as a result of sloppy programming or poor developer discipline.
Instead, they consider it as a result of intentional decisions to trade off competing concerns during development \citep{klinger2011enterprise}.
More specifically, such technical debt is the comment that is \textbf{intentionally} introduced by developers to alert the inadequacy of the solution \citep{Potdar_ICSME2014} and is acknowledged by developers.
For example, in the open-source project TensorFlow, a comment saying \textit{TODO(b/26910386): Identify why this infrequently causes timeouts.}, indicates that the corresponding code is problematic and needs further investigation.

In this paper, we use the introduction and removal of SATD instances in a family of software systems as an example, e.g., the development of deep learning frameworks, to find the common patterns on how developers remove different types of technical debt in this software family.
We employ the development of deep learning frameworks as an example because we would like to choose a homogeneous set of projects from the same domain to minimize the domain confounding effect.
The field of the development of different deep learning frameworks is the same, i.e., offering high-level programming interfaces to deep learning applications by the implementation of a range of concrete tasks, e.g., implementing core building blocks for designing, training, and validating deep neural networks.
Moreover, deep learning frameworks are arguably some of the most important software systems today, due to the wide use of deep learning applications and its prevalence in health \citep{litjens2017survey}, cars \citep{sallab2017deep, huval2015empirical, al2017deep, shalev2016safe}, etc, i.e., life-impacting software.
Hence, the effective and efficient maintainability of deep learning frameworks is of critical importance.
However, the deep learning related techniques are still rapidly advancing, with many cutting edge technologies being continuously proposed, which cover a wide range of knowledge, e.g., Generative Adversarial Networks (GAN)\footnote{https://www.tensorflow.org/api\_docs/python/tf/contrib/gan} \citep{gan_2014}, Tensor Processing Unit (TPU)\footnote{https://www.tensorflow.org/api\_docs/python/tf/contrib/tpu} \citep{jouppi2017datacenter}, and Batch Normalization\footnote{https://mxnet.incubator.apache.org/api/python/symbol/symbo-l.html\#mxnet.symbol.BatchNorm} \citep{ioffe_batch_2015}.
Developers have to implement these novel techniques in time to win the fierce market competition, which increases the risk of technical debt at the same time.
As a part of the effective and efficient maintainability of deep learning frameworks, the removal of technical debt is a crucial aspect.

To do so, we first extract all the comments in all versions of files, and then we identify the SATD instances by SATD-detector \citep{liu2018satd, huang2018identifying}.
When we started our research, SATD-detector \citep{liu2018satd, huang2018identifying} was the most advanced NLP based algorithm to automatically identify the SATD instances.
We re-train the SATD-detector with the comments that are presented in  \citet{liu_icse2020}'s work, where they manually labeled the comments in the latest stable version of the deep learning frameworks we studied.
Finally, based on \citet{liu_icse2020}'s work, we manually label the detected SATD instances into different types by card-sorting \citep{spencer2009card}.
We identify the 7 types of technical debts that are the same as \citet{liu_icse2020}'s work: design debt (i.e., sub-optimal design), documentation debt (i.e., incomplete documentation), defect debt (i.e., unresolved defects), requirement debt (i.e., incomplete implementation of the methods), test debt (i.e., deficiencies in tests), algorithm debt (i.e., sub-optimal algorithm), and compatibility debt (i.e., immature dependencies).
We observe that 75\% of SATD instances that are introduced before the latest stable version are removed in 299 days at the most (for PyTorch).
To avoid the bias caused by the SATD instances that are introduced recently before the latest stable version (i.e., right censoring) \citep{quesenberry1989survival},
\textbf{we investigate the introduction of different types of technical debt instances that are introduced over one year before the latest stable release version.
Then, we characterize their removal before the latest stable release version}.
More specifically, with the data, we characterize the removal of technical debt by exploring several questions:

\noindent(1) \textbf{Which types of technical debt are prevalently introduced along the development process?}\\
The distribution of different types of technical debt that are introduced along the development process can reflect what kind of sub-optimal trade-offs or decisions are made by developers during the development process.
We find that developers introduce the design debt the most during the development, followed by requirement debt and algorithm debt.

\noindent(2) \textbf{Which types of technical debt are removed the most?}\\
The distribution of different types of technical debt among the removed SATD instances along the development process can reflect the allocation of developers' effort in the resolution of technical debt.
The proportion of introduced technical debt instances that are removed along the development process can reflect which types of technical debt are removed the most to finish the development tasks and ensure the code quality.
We find that requirement debt is removed the most, followed by design debt.

\noindent(3) \textbf{Which types of technical debt are removed the fastest?}\\
The survival time of the technical debt, i.e., the time interval since the introduction to the removal of technical debt, can illustrate the priority of different types of technical debt when developers resolve them.
We find that design debt is removed the fastest, followed by requirement debt.

\noindent(4) \textbf{Who removes different types of technical debt?}\\
It is expected that the technical debt is self-removed, i.e., removed by the developer who introduces it.
As a result, most of test debt, design debt, and requirement debt are removed by the developers who introduced them.

Based on the introduction and removal of different types of technical debt instances, we depict the evolution of the frequencies of different types of technical debt that are presented in different stages.
This evolution can illustrate the changes in developers' concerns.
We also discuss the removal patterns of different types of technical debt, highlight future research directions, and provide recommendations for practitioners.

\vspace{0.1cm}\noindent{\bf Paper Organization.}
The remainder of this paper is structured as follows.
Section \ref{setup} presents the detailed approaches we use to collect and pre-process data.
Section \ref{findings} presents our research findings.
Section \ref{discussion} depicts the evolution of the frequencies of different types of technical debt along the development process, summarizes the removal patterns of different types of technical debt, presents the implications and actionable suggestions based on our findings, and describes some threats to the validity of this study.
Section \ref{related} presents the related work of our study, including the research works on technical debt and research works on software engineering for deep learning. We also compare the findings in our work with the findings in previous work.
Finally, Section \ref{conclusion} concludes our study and presents future work.

\section{Case Study Setup}\label{setup}
In this section, we describe the steps that we took for project selection, project data extraction, source code comments extraction, SATD instances identification and manual classification.

\subsection{Project Selection}

We focus on open-source deep learning frameworks hosted in Github. We exclude deep learning applications that build upon such frameworks, or general-purpose mathematical libraries that those deep learning frameworks build upon.
To do so, we first search repositories labeled by \emph{deeplearning} and \emph{deep learning} topics\footnote{https://blog.github.com/2017-01-31-introducing-topics/} in GitHub.
Then, we identify the deep learning framework projects by reading the readme file of the projects.
As a result, we include 7 deep learning frameworks with the largest number of stars that are written in 3 programming languages (C++, Python and Java) as subject frameworks for our study. They include: TensorFlow (shortened as TF)\footnote{https://github.com/tensorflow/tensorflow}, Keras\footnote{https://github.com/keras-team/keras}, Deeplearning4j (shortened as DL4J)\footnote{https://github.com/deeplearning4j/deeplearning4j}, Caffe\footnote{https://github.com/BVLC/caffe}, PyTorch\footnote{https://github.com/pytorch/pytorch}, MXNet\footnote{https://github.com/apache/incubator-mxnet} and Microsoft Cognitive Toolkit (known as CNTK)\footnote{https://github.com/Microsoft/CNTK}.

Table \ref{table_dataset} provides statistics of each framework in the latest stable version in our study, including the release version, the total number of lines of code, the total number of commits, the number of contributors and the main programming languages.
Following the previous study by \citet{Maldonado_MTD2015}, we calculate the total number of code lines using SLOCCount\footnote{https://dwheeler.com/sloccount/}.

\begin{table}
\centering
\caption{Overview of Studied Projects}\label{table_dataset}
\resizebox{\linewidth}{!}{
\begin{tabular}{c|ccccc}
\toprule
\textbf{Framework}      & \textbf{Release} & \textbf{\#Lines of Code} & \textbf{\# Commits} & \textbf{\#Contributors} & \textbf{Languages}   \\
\midrule
TF             & v1.9.0  & 1,821,016    & 34,227     & 1,868         & Python, C++ \\
Keras          & 2.2.2   & 42,182       & 4,651      & 782           & Python      \\
Caffe          & 1.0.0   & 76,322       & 4,020      & 318           & C++         \\
PyTorch        & v0.4.0  & 617,255      & 10,835     & 1010          & Python, C++ \\
MXNet          & 1.2.1   & 305,755      & 7,015      & 682           & Python, C++ \\
CNTK           & v2.5.1  & 324,472      & 15,575     & 269           & Python, C++ \\
DL4J           & 0.9.1   & 361,366      & 8,375      & 185           & Java        \\  \bottomrule
\end{tabular}
}
\end{table}

\subsection{Comment Extraction}\label{comment_extraction}

We need the comments in all of the file history of the selected deep learning frameworks.
We follow the steps performed in Maldonado et al.'s work \citep{maldonado_icsme2017}.
More specifically, since we are interested in when the SATD is removed and who removed the SATD during the whole development process, we investigate the introduction and removal of SATD along with the commits on the master branch.

To do so, we first obtain all the versions of files along the master branch.
We identify all Java, C++, and Python source code files currently available in the latest version of the project, and then we check out each version of the repository to get deleted files in each commit, which are currently absent in the latest stable version.
We identify the rename or move of the file with Git.
Finally, we obtain all versions of files by tracking all the commits done to each file.

After obtaining all versions of files in the software repositories, we discriminate between source code and comment lines. 
We use the srcML Toolkit\footnote{https://www.srcml.org/}, which is capable of parsing source files that are coded by C++ and Java, into XML files.
For Python files that are not supported by srcML, we utilize the tokenize module\footnote{https://docs.python.org/3/library/tokenize.html} in the Python standard library, which provides a lexical scanner for Python source code to identify all comments.
We record the file name, the class name, the method name, and the comment content, as well as the meta-information of the version of the file that contains the comments, e.g., the creation date, the creation user email, the commit id, which is obtained from the version control system, i.e., Git.

To track the introduction and removal of the comment, we consider the first available file version that contains the comment as the file version that introduced the comment.
Similarly, we consider the first version that the comment instance does not exist or the file where the comment exists is deleted as the removal version.
The file where the comment exists being deleted also indicates that the comment does not exist.
In certain cases, a comment is found in one version only (i.e., the version that it is introduced in), which indicates a scenario where the comment is introduced and removed immediately after the introduction.
Moreover, there can be inconsistent changes between the comments and the code, i.e., in some cases the comment may change but not the code, and vice versa, we will discuss this threat in Section \ref{threats}.
Finally, we extract a total of 445,149 distinct comments in all versions of the files.

\subsection{Identification of SATD Instance}\label{satdextraction}

To identify technical debt, we follow \citeauthor{maldonado_icsme2017}'s work, which uses an NLP based algorithm to automatically identify the comments that indicate technical debt.
When we started our research, SATD-detector \citep{liu2018satd, huang2018identifying} was the most advanced NLP based algorithm to automatically identify the SATD instances.
More specifically, to build the model, SATD-detector preprocesses the text descriptions of comments and extracts features (i.e. words) to represent each comment at first.
Then Information Gain (IG) is employed to select features that are useful for classification and remove useless features.
Finally, the selected features are used to train a classifier for each project.
In the prediction phase, the comment is processed to extract features.
Then the features are inputted to the trained composite classifier.
Finally, each sub-classifier will predict the label of the comment according to its features, and the label with the largest number of ``votes'' will be chosen as the final prediction result of the composite classifier.

To ensure the accuracy of SATD-detector in detecting \textbf{the SATD instances in all versions of files} in deep learning frameworks, \textbf{we re-train the SATD-detector with the comments of the deep learning frameworks which are labeled in \citep{liu_icse2020}'s work.}
In \citep{liu_icse2020}'s work, they manually label the comments in the latest stable version of the 7 deep learning frameworks as we studied and find that there are 7,159 SATD instances.
This indicates that there is an overlap between the training dataset and the test dataset.
This could lead to higher performance for SATD-detector in identifying the SATD instances in all versions of files in deep learning frameworks.
We report the precision and recall of the SATD-detector after labeling the SATD instances in the identified comments in Section \ref{manually_classify}.

\subsection{Manual Classification}\label{manually_classify}
To determine the different SATD types, we utilize the SATD types which are found in \citet{liu_icse2020}'s work as a starting point, where they analyze the comments of 7 popular deep learning frameworks as we studied.
They find that the technical debt in deep learning frameworks can be classified into seven categories: design debt, defect debt, documentation debt, requirement debt, test debt, algorithm debt, and compatibility debt.

In our paper, we perform two iterations of a card sorting approach \citep{spencer2009card} to classify 29,778 detected SATD instances in these 7 deep learning frameworks.
Concretely, in the first iteration of classification, we try to ensure that our classification standard is consistent with previous work.
To do so, we first randomly pick 100 comments from the dataset provided by \citet{liu_icse2020}'s work, then the first two authors manually classify these sentences according to \citet{liu_icse2020}'s work.
A discussion on the disagreements with \citet{liu_icse2020}'s work is performed after the classification process.
To validate our classification standard, the first author selected another 500 comments from the dataset provided by \citet{liu_icse2020}'s work and manually classified them.
Then, we calculate the Cohen's kappa coefficient \citep{cohen_kappa} and obtain a result of +0.85, which indicates a high level of agreement with the classification given by the first author and \citet{liu_icse2020}'s work.
During this phase, the coding schema of different types of technical debt in deep learning frameworks is revised.

In the second classification iteration, the first author classifies all the 29,778 detected SATD instances.
During this phase, the categories of the SATD instances in all versions of files are identified.
To reduce personal bias in the manual classification of code comments, we randomly sampled a statistically representative sample of 1,000 SATD instances from the 29,778 detected SATD instances using a 95\% confidence level with a 10\% confidence interval.
We invite an independent Ph.D. student, who is not an author of this paper, to manually classify the randomly sampled 1,000 SATD instances.
We discuss the disagreements in Section \ref{threats}.
A high level of agreement between the classification given by the two different students is reported with Cohen's kappa coefficient of +0.79. This gives us high confidence in the dataset used in our paper.

As a result, we find that there are 24,032 SATD instances in all versions of files in deep learning frameworks.
We observe that 75\% of the SATD instances that are introduced before the latest stable version are removed in 299 days at the most (for PyTorch).
To avoid bias caused by the SATD instances that are introduced recently before the latest stable version (i.e., right censoring) \citep{quesenberry1989survival}, \textbf{we exclude the SATD instances that are introduced in one year before the latest stable release version.}
More specifically, \textbf{we investigate the introduction of different types of technical debt instances that are introduced over one year before the latest stable release version.
Then, we characterize their removal before the latest stable release version.}

We discuss the performance of SATD-detector in identifying the SATD instances that are studied in our work.
For 318,044 comments that are introduced over one year before the latest stable release version, 21,702 of them are identified as SATD instances by SATD-detector, and 17,576 of them are classified as SATD instances by our manual classification process.
This shows that the retrained SATD-detector achieves a precision score of 0.81 as 4,126 comments are false positive (i.e., not the comments indicating technical debt).
Table \ref{table_introduced_types} reports the precision scores of SATD-detector in identifying the SATD instances in different projects.
Besides, we observe that the false positive instances are almost uniformly introduced in different years (normalized entropy scores range from 0.76 for PyTorch to 0.97 for Keras).
Therefore, moving further back into the evolutionary history of the project would not affect the performance of the SATD-detector in identifying SATD instances.
To check the recall of SATD-detector in identifying the comments indicating technical debt, we randomly sampled a statistically representative sample of 100 comments from 318,044 comments using a 95\% confidence level with a 10\% confidence interval.
We find that there is only 1 comment indicating technical debt.
This shows that the retrained SATD-detector achieves a recall score of 0.85 as there are around 3,180 comments that are false negative (i.e., the comments indicating technical debt). 
We identify the following types of technical debt, which are the same as Liu et al.'s work:

\noindent (1) \textbf{Design debt} indicates sub-optimal design, e.g., misplaced code, lack of abstraction, long methods, poor implementation, workarounds, or temporary solutions on the usage of other internal functions. 

\noindent Example: 
\emph{``TODO(b/32239616): This kernel should be moved into Eigen and vectorized.''} - [TF]\footnote{tensorflow/tensorflow/core/kernels/cwise\_ops.h}

\noindent (2) \textbf{Defect debt} corresponds to code that behaves in unintended ways, and developers postpone repairing it because of various factors (e.g., time-consuming to resolve).

\noindent Example: 
\textit{``Linear weights do not follow the column name. But this is a rare use case, and fixing it would add too much complexity to the code.''} - [TF]\footnote{tensorflow/tensorflow/python/feature\_column/feature\_column\_test.py}

\noindent (3) \textbf{Documentation debt} indicates missing, inadequate or incomplete documentation that explains the corresponding part of the program.

\noindent Example: 
\textit{``TODO(sibyl-vie3Poto): Write up a doc with concrete derivation and point to it from here."} - [TF]\footnote{tensorflow/tensorflow/core/kernels/hinge-loss.h}

\noindent (4) \textbf{Requirement debt} indicates \emph{incompleteness} of the method, class or program at the time, which may mean that the original planned completion of the task exceeds the development schedule. It can also correspond to cases when new requirements are identified during the development of existing requirements but cannot be considered due to time pressure or other constraints.

\noindent Example: 
\textit{``TODO setup for RNN''} - [DL4J]\footnote{deeplearning4j/deeplearning4j-nn/src/main/java/org/deeplearning4j/nn/params/Batch-NormalizationParamInitializer.java}

\noindent (5) \textbf{Test debt} indicates the need for improvements to address deficiencies in the test suite.

\noindent Example: 
\textit{``TODO(fchollet): insufficiently tested.''} - [TF]\footnote{tensorflow/tensorflow/python/keras/backend\_test.py}

\noindent (6) \textbf{Compatibility debt} refers to the debt related to a project's immature dependencies on other projects, which cannot supply all qualified services, and the current implementation is a temporary workaround.

\noindent Example: 
\textit{``Moved to common.cpp instead of including boost/thread.hpp to avoid a boost/NVCC issues (\#1009, \#1010) on OSX. Also fails on Linux with CUDA 7.0.18.''} - [Caffe]\footnote{caffe/include/caffe/common.hpp}

\noindent (7) \textbf{Algorithm debt} refers to the debt that the algorithm implemented in a deep learning framework is sub-optimal.

\noindent Example: 
\textit{``TODO(Yangqing): Is there a faster way to do pooling in the channel-first case?''} - [Caffe]\footnote{caffe/src/caffe/layers/pooling\_layer.cpp}\\

\section{Findings}\label{findings}
In this section, we first investigate the distribution of the introduced technical debt, and then we quantify the removal of technical debt from different perspectives, such as removal rate, removal pace, and self-removal rate.

\subsection{RQ1: Which types of technical debt are prevalently introduced along the development process?}\label{introduced}

\vspace{0.1cm}\noindent\textbf{Motivation:}
In \citep{liu_icse2020}'s work, they observed different types of technical debt are in the latest stable version of 7 deep learning frameworks.
However, it is still unclear which types of technical debt are prevalently introduced along the development process.
The admitted technical debt indicates the acknowledgment of sub-optimal trade-offs or decisions during the development process.
To summarize which types of sub-optimal trade-offs or decisions developers would admit more during the development process, we investigate the distribution of different types of technical debts introduced throughout the development process.

\vspace{0.1cm}\noindent\textbf{Approach:}
To better describe the introduction of different types of technical debt along the development process, we first analyze the distribution of different types of technical debt that are introduced along the development process, then we present the distribution of different types of technical debt of all technical debt instances that are introduced one year before the latest stable version. 

\begin{figure}
\centering
\includegraphics[width = \linewidth]{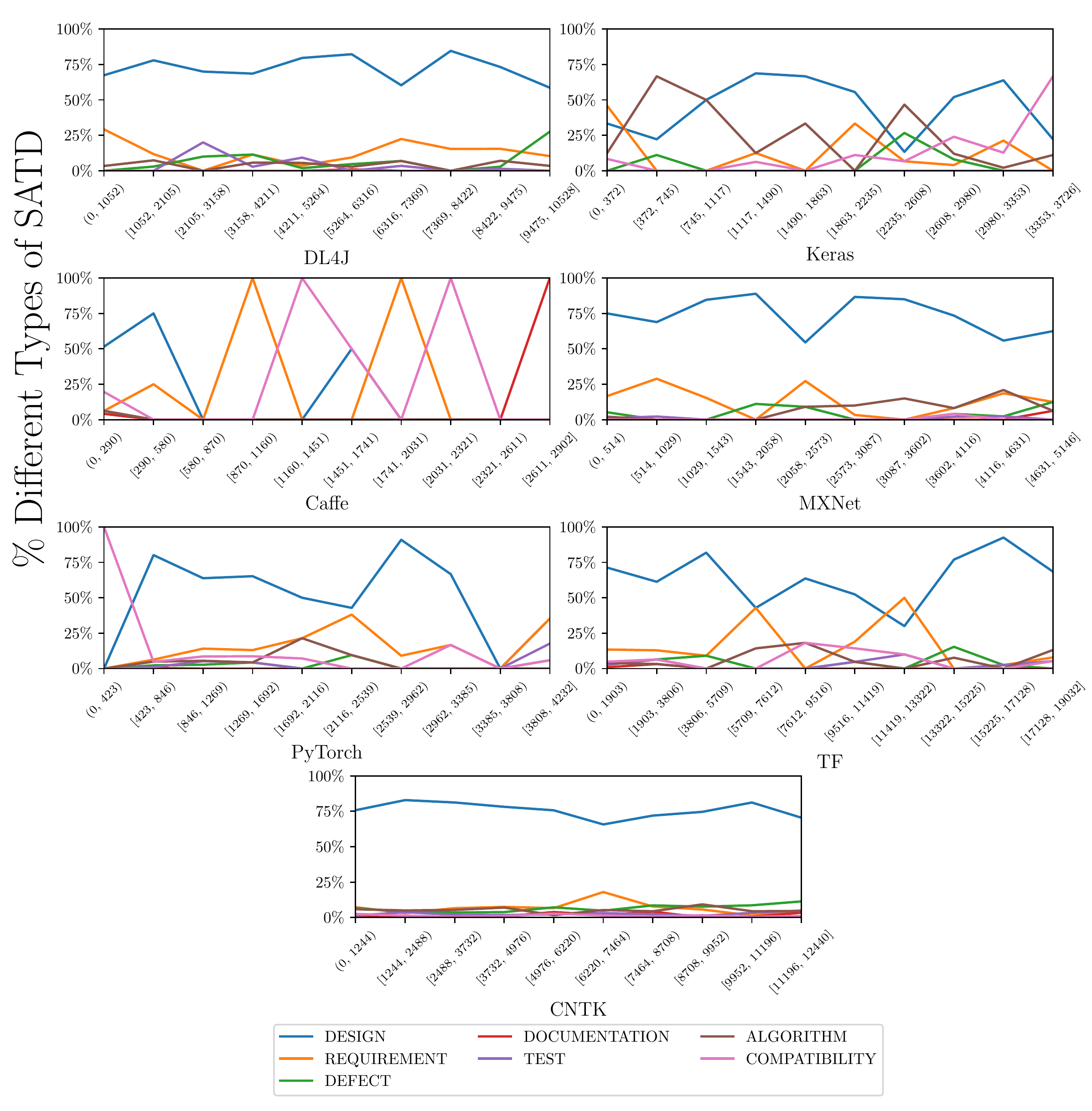}
\caption{Distribution of different types of SATD instances that are introduced in different development phases.
}\label{figure_introduction_formation}
\end{figure}

To characterize the distribution of different types of technical debt along the development process, we first divide the whole development process one year before the latest stable release version into ten \textbf{development phases} based on the chronological order of the commits.
Then we count the number of different types of technical debt instances that are introduced in each development phase.
Since different numbers of SATD instances are introduced in different development phases, we normalize different types of SATD instances by the number of total SATD instances that are introduced in that development phase.
For example, in TensorFlow, there are 19,032 commits along the development process one year before the latest stable release version.
We first divide the whole development process one year before the latest stable release version into 10 development phases.
Each development phase has 1,903 commits.
Then, we count the number of different types of SATD instances that are introduced in each development phase and normalize them with the total number of the SATD instances introduced in that phase.
Figure \ref{figure_introduction_formation} shows the distributions of different types of the introduced SATD instances along the developing process in 7 deep learning frameworks.

To check whether the difference between different types of technical debt in terms of their proportions among the SATD instances that are introduced along the development process is statistically significant, we perform a Kruskal-Wallis H test \citep{kruskal1952use}.
Kruskal-Wallis H test is a non-parametric test for comparing whether two or more independent samples originate from the same distribution.
As a result, we find that \textbf{the difference between different types of technical debt in terms of its proportions in introduced SATD instances along the development process is significant (p-value $<$ 0.05).}
Then, we perform a Dunn's test with Bonferroni correction to determine which groups differ from each other group \citep{dunn1961multiple}.
Dunn's test can be used for the post-hoc analysis for the specific sample pairs.
To calculate the effect size, we calculate the corresponding Cliff's deltas \citep{cliff1993dominance}.
Cliff's delta is a measure of how often the values in one distribution are larger than the values in a second distribution.
Table \ref{table_introduced_test} presents p-values and Cliff's deltas.

\begin{table}[!htb]
\centering
\caption{P-values and Cliff's deltas of the differences between different types of technical debt in terms of their proportions among the SATD instances that are introduced along the development process.
We report the pairs of SATD types with p-value $<$ 0.05 (i.e., significant) with $\star$ and the interpretation of corresponding Cliff's deltas.
}\label{table_introduced_test}
\footnotesize
\resizebox{\linewidth}{!}{
\begin{tabular}{c|cccccc}
       \toprule
\textbf{}     & \textbf{Algorithm}  & \textbf{Compatibility} & \textbf{Defect}     & \textbf{Design}    & \textbf{Documentation} & \textbf{Requirement} \\
\hline
Compatibility &                     &                        &                     &                    &                        &                      \\
Defect        &                     &                        &                     &                    &                        &                      \\
Design        & $\star$   (+large)  & $\star$   (+large)     & $\star$   (+large)  &                    &                        &                      \\
Documentation & $\star$   (-large)  & $\star$   (-medium)    & $\star$   (-medium) & $\star$   (-large) &                        &                      \\
Requirement   &                     & $\star$   (+medium)    & $\star$   (+medium) & $\star$   (-large) & $\star$   (+large)     &                      \\
Test          & $\star$   (-medium) &                        &                     & $\star$   (-large) &                        & $\star$   (-large)  \\
\bottomrule
\end{tabular}
}
\end{table}

Table \ref{table_introduced_types} presents an overview of the distribution of different types of technical debt in different frameworks, as well as the total number of SATD instances that are introduced one year before the latest stable release version for each project.
To better view the differences between different types of technical debt, we highlight the top three types in terms of the proportion in each project in bold.

\begin{table}[!htb]
\centering
\caption{Distribution of different types of SATD that are introduced one year before the latest stable release version.
We also report the precision scores of the retrained SATD-detector in detecting to SATD instances in different deep learning frameworks.
}\label{table_introduced_types}
\resizebox{\linewidth}{!}{
       \begin{tabular}{c|cccccccc}
       \toprule
       \textbf{Project   Name} & \textbf{TF}      & \textbf{Keras}   & \textbf{CNTK}    & \textbf{Caffe}   & \textbf{MXNet}   & \textbf{PyTorch} & \textbf{DL4J}    & \textbf{Average} \\
       \hline
       \textbf{Total}          & 5622             & 191              & 8398             & 270              & 432              & 1577             & 581              & 2438.7           \\
       \hline
       \textbf{Design}         & \textbf{65.71\%} & \textbf{51.31\%} & \textbf{76.33\%} & \textbf{48.52\%} & \textbf{74.31\%} & \textbf{68.80\%} & \textbf{72.46\%} & \textbf{65.35\%} \\
       \textbf{Compatibility}  & 2.81\%           & 12.04\%          & 1.70\%           & \textbf{10.00\%} & 1.39\%           & 5.64\%           & 0.00\%           & 4.80\%           \\
       \textbf{Defect}         & 3.33\%           & 5.24\%           & \textbf{5.54\%}  & 4.07\%           & 3.01\%           & 3.04\%           & \textbf{5.34\%}  & 4.22\%           \\
       \textbf{Documentation}  & 1.03\%           & 0.00\%           & 1.06\%           & \textbf{20.00\%} & 0.23\%           & 0.25\%           & 0.17\%           & 3.25\%           \\
       \textbf{Test}           & 5.12\%           & 0.52\%           & 2.66\%           & 2.59\%           & 1.16\%           & 2.85\%           & 2.58\%           & 2.50\%           \\
       \textbf{Algorithm}      & \textbf{6.05\%}  & \textbf{14.14\%} & 5.51\%           & 7.41\%           & \textbf{5.79\%}  & \textbf{5.64\%}  & 4.82\%           & \textbf{7.05\%}  \\
       \textbf{Requirement}    & \textbf{15.96\%} & \textbf{16.75\%} & \textbf{7.04\%}  & 7.41\%           & \textbf{14.12\%} & \textbf{13.76\%} & \textbf{14.63\%} & \textbf{12.81\%} \\
       \hline
       \textbf{Precision} & 0.71 & 0.45 & 0.91 & 0.79 & 0.60 & 0.83 & 0.81 & 0.81  \\
       \bottomrule
       \end{tabular}
}
\end{table}

\vspace{0.1cm}\noindent\textbf{Results:}
Table \ref{table_introduced_test} shows that the differences between design debt and other types of technical debt are significantly and large.
Figure \ref{figure_introduction_formation} shows that design debt is the most introduced technical debt along the development process with fluctuation in MXNet, CNTK, and DL4J.
Design debt is the most introduced technical debt in most of the development phases in TensorFlow and Keras.
This shows that \textbf{design debt is the most common technical debt across deep learning frameworks along the development process}.
During the development process, developers are not satisfied with the design of the code.
Developers admit the inadequacy of the design of the code the most.
Caffe is an outlier here, design debt is the most introduced technical debt at the beginning of the development process, and following that, either requirement debt or test debt is the most introduced technical debt.
One possible reason is that the Caffe planned to migrate to a new project, i.e., Caffe2.
Therefore, in the later phase of development, the developers in Caffe pay attention more to the implementation of requirements and the completeness of test cases, rather than the design of code.
Finally, until one year before the latest stable release version of 7 deep learning frameworks, the proportions of the introduced design debt among all introduced technical debt instances range from 76.33\% in CNTK to 48.52\% in Caffe.

Table \ref{table_introduced_test} shows that the differences between requirement debt and other types of technical debt except algorithm debt are significant, and the effect sizes range from medium to large.
Figure \ref{figure_introduction_formation} shows that requirement debt is one of the second most commonly introduced technical debt along the time in 7 deep learning frameworks, e.g., in PyTorch.
This shows that \textbf{requirement debt is the second most common technical debt}.
Developers frankly wrote down the unaccomplished tasks in comments as a notification during the development process.
In certain development phases, e.g., the 7th development phase in TensorFlow, the 6th development phase in PyTorch, requirement debt is introduced more.
One possible reason is that developers are arranged to finish more requirements that exceed their ability.
The unaccomplished requirements are left as requirement debt.
Finally, until one year before the latest stable release version of 7 deep learning frameworks, the proportions of requirement debt instances among all introduced technical debt in different frameworks range from 7.04\% in Caffe to 16.75\% in Keras.

Table \ref{table_introduced_test} shows that algorithm debt is significantly different from documentation debt and test debt, and the effect sizes range from medium to large.
Figure \ref{figure_introduction_formation} shows that different from the design debt and requirement debt that are introduced more along the development process, algorithm debt is introduced more in certain development phases.
This shows that \textbf{algorithm debt is the third most common technical debt}.
For example, in MXNet, design debt, requirement debt, and defect debt is introduced more before the 5th development phases.
Since the 5th development phase, more than 10\% of the introduced technical debt is algorithm debt.
One possible reason is that developers transfer their attention from the design of code and the implementation of functions to the optimization of algorithm.
Finally, until one year before the latest stable release version of 7 deep learning frameworks, the proportions of algorithm debt in different frameworks range from 4.82\% in DL4J to 14.14\% in Keras.

Table \ref{table_introduced_test} shows that documentation debt is significantly different from other types of technical debt except for test debt, and the effect size range from medium to large.
Along the development process, we hard to observe the introduction of documentation debt.
This shows that \textbf{documentation debt is the least common debt}.
Considering quantities of comments and documentation of the deep learning frameworks, the documentation debt can be related to that developers seldom perform sub-optimal trade-offs or decisions related to documentation.
Finally, until one year before the latest stable release version of 7 deep learning frameworks, the proportions of documentation debt in different frameworks range from 0\% in Keras to 20.0\% in Caffe.

Table \ref{table_introduced_test} shows that test debt is significantly different from algorithm debt, design debt, and requirement debt, and the effect size range from medium to large.
This shows that \textbf{test debt is the second least introduced technical debt}.
The small proportions of test debt do not mean that there are fewer sub-optimal trade-offs or decisions related to the testing in the projects.
One possible reason is the wide use of professional test management systems, such as QTest\footnote{https://www.qasymphony.com/software-testing-tools/qtest-manager/test-case-management/}.

\vspace{0.2cm}\noindent \fbox{\parbox[c]{1\linewidth} {\emph{Design debt is the most prevalent technical debt along the development process, followed by requirement debt and algorithm debt.
Documentation debt is the least common technical debt along the development process, followed by test debt.
}}}

\subsection{RQ2: Which types of technical debt are removed the most?}\label{remove_rate}

\vspace{0.1cm}\noindent\textbf{Motivation:}
In this section, we would like to characterize the removal of different types of technical debt in deep learning frameworks along the development process.
In Section \ref{introduced}, we observe the prevalence of different types of technical debt in deep learning frameworks.
To ensure the code quality, developers are expected to take actions (e.g., factoring) to resolve these SATD instances.
However, it is still unclear which types of technical debt is removed the most along the development process.

\vspace{0.1cm}\noindent\textbf{Approach:}
To better describe the removal of different types of technical debt along the development process, we first analyze the distribution of different types of technical debt that are removed along the development process.
By doing so, we could understand which types of technical debt attract development attention more in different development phases.
Then we investigate the proportion of different types of introduced SATD instances that are removed along the development process.
By doing so, we could understand which types of technical debt are removed the most to finish the development tasks and ensure the code quality.
Finally, we present the proportion of different types of technical debt of all technical debt instances that are removed before the latest stable version. 

\begin{figure}
\centering
\includegraphics[width = \linewidth]{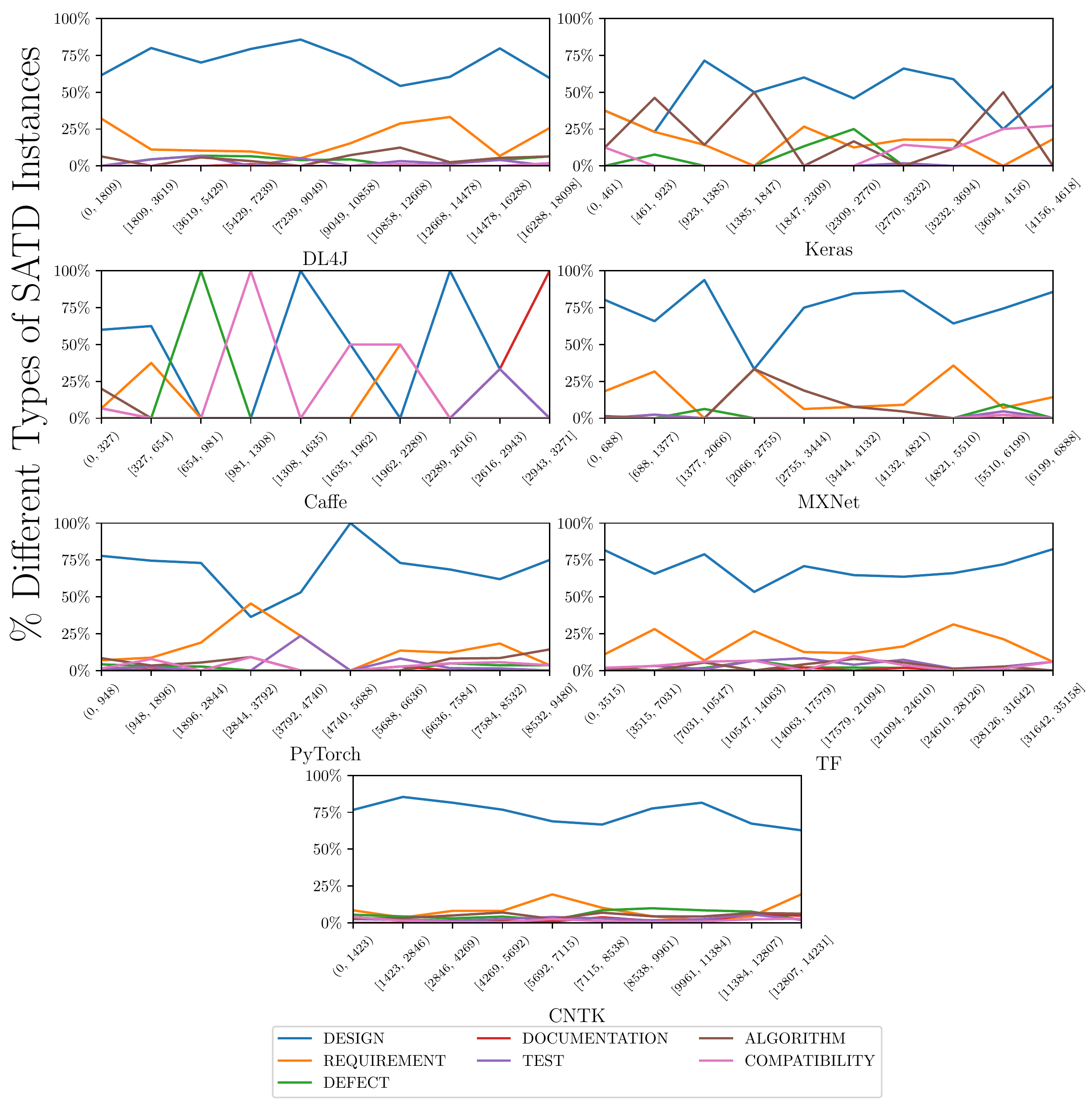}
\caption{Distribution of different types of technical debt that are removed in different development phases.}\label{figure_removal_formation}
\end{figure}

To describe the removal of different types of technical debt along the development process, we divide the whole development process into ten \textbf{development phases} based on the chronological order of the commits.
We first analyze the distribution of different types of technical debt among the technical debt that is removed in different development phases.
More specifically, we normalize the number of different types of technical debt that are removed in each development phase by the total number of technical debt instances that are removed in that development phase.
Figure \ref{figure_removal_formation} shows the distribution of different types of technical debt among the removed technical debt instances in different development phases.

\begin{table}[!htb]
\centering
\caption{P-values and Cliff's deltas of the differences between different types of technical debt in terms of their proportions among the removed technical debt instances along the development process.
We report the pairs of SATD types with p-value $<$ 0.05 (i.e., significant) with $\star$ and present the interpretation of the corresponding Cliff's delta.
}\label{table_removal_prop_test}
\resizebox{\linewidth}{!}{
\begin{tabular}{c|cccccc}
\toprule
\textbf{}     & \textbf{Algorithm} & \textbf{Compatibility} & \textbf{Defect}   & \textbf{Design}  & \textbf{Documentation} & \textbf{Requirement} \\ \hline
Compatibility &                    &                        &                   &                  &                        &                      \\
Defect        &                    &                        &                   &                  &                        &                      \\
Design        & $\star$ (+large)   & $\star$ (+large)       & $\star$ (+large)  &                  &                        &                      \\
Documentation & $\star$ (-large)   & $\star$ (-medium)      & $\star$ (-medium) & $\star$ (-large) &                        &                      \\
Requirement   & $\star$ (+large)   & $\star$ (+large)       & $\star$ (+large)  & $\star$ (-large) & $\star$ (+large)       &                      \\
Test          & $\star$ (-medium)  &                        &                   & $\star$ (-large) & $\star$ (+medium)      & $\star$ (-large)     \\ \bottomrule
\end{tabular}
}
\end{table}

To check whether the differences between different types of technical debt in terms of their proportions among the removed technical debt instances along the development process are significant, we perform a Kruskal-Wallis H test \citep{kruskal1952use}.
Kruskal-Wallis H test is a non-parametric test for comparing whether two or more independent samples originate from the same distribution.
As a result, we find that \textbf{the differences between different types of technical debt in terms of their proportions in removed technical debt instances are significant (p-value $<$ 0.05).}
Then, we perform a Dunn's test with Bonferroni correction to determine which groups differ from each other group \citep{dunn1961multiple}.
Dunn's test can be used for the post-hoc analysis for the specific sample pairs.
To calculate the effect size, we calculate the corresponding Cliff's deltas \citep{cliff1993dominance}.
Cliff's delta is a measure of how often the values in one distribution are larger than the values in a second distribution.
Table \ref{table_removal_prop_test} presents p-values and Cliff's deltas.

\begin{figure}
\centering
\includegraphics[width = \linewidth]{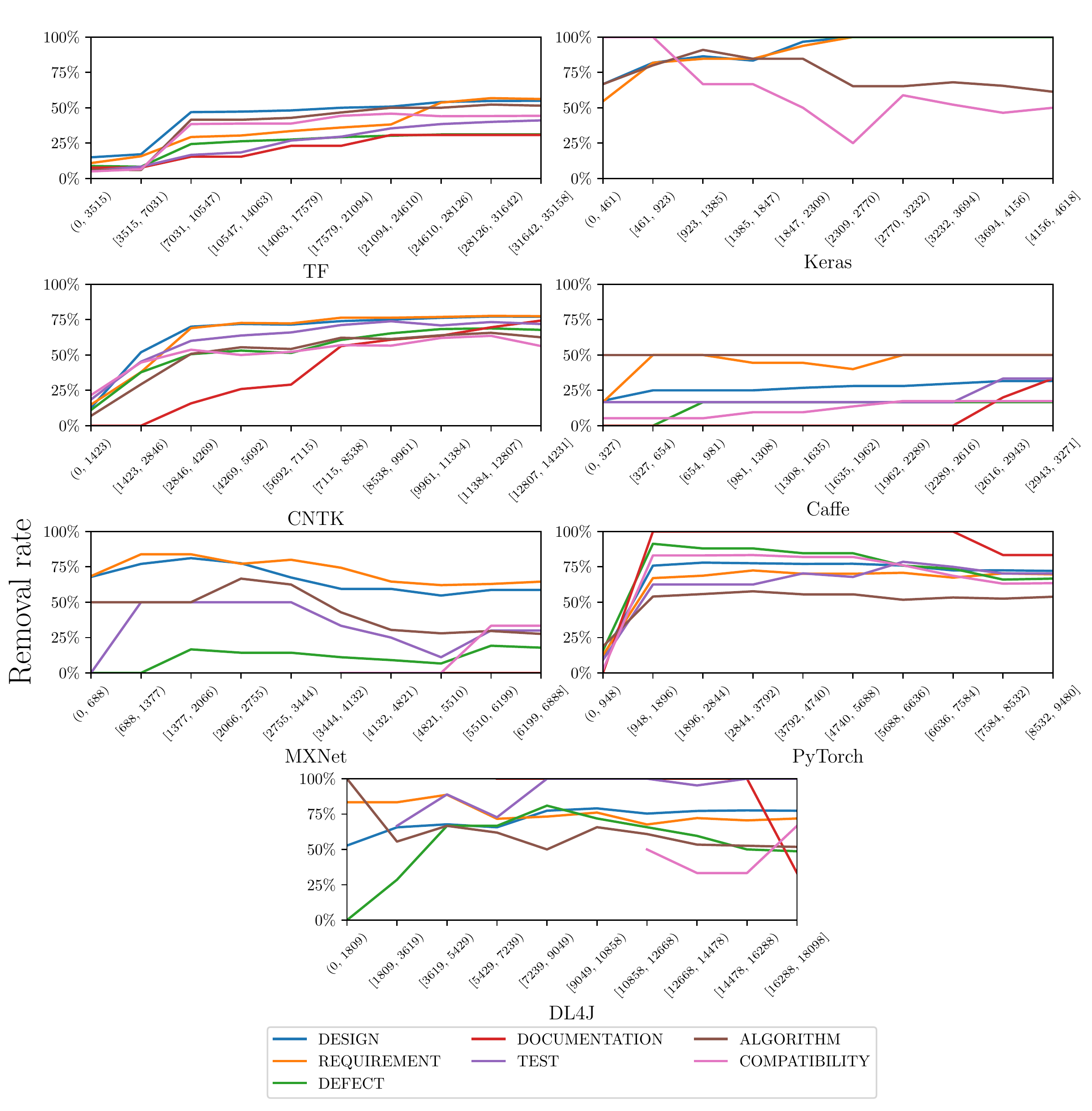}
\caption{Removal rate of different types of technical debt along the development process.}\label{figure_removal_rate_format}
\end{figure}

Then we analyze the removal rate (i.e., the proportion of removed SATD instances among the introduced SATD instances) of different types of technical debt instances at different development phases.
For certain development phases, the \textbf{removal rate} of different types of technical debt instances is calculated as \textbf{the proportion of the removed technical debt among the introduced different types of technical debt instances.}
Figure \ref{figure_removal_rate_format} shows the removal rate of the existing technical debt instances.

To check whether the differences between different types of technical debt in terms of their removal rates among the introduced technical debt instances along the development process are significant, we perform a Kruskal-Wallis H test \citep{kruskal1952use}.
Kruskal-Wallis H test is a non-parametric test for comparing whether two or more independent samples originate from the same distribution.
As a result, we find that \textbf{the differences between different types of technical debt in terms of their removal rates in introduced technical debt instances are significant (p-value $<$ 0.05).}
Then, we perform a Dunn's test with Bonferroni correction to determine which groups differ from each other group \citep{dunn1961multiple}.
Dunn's test can be used for the post-hoc analysis for the specific sample pairs.
To calculate the effect size, we calculate the corresponding Cliff's deltas \citep{cliff1993dominance}.
Cliff's delta is a measure of how often the values in one distribution are larger than the values in a second distribution.
Table \ref{table_removal_rate_test} presents p-values and Cliff's deltas.

\begin{table}[!htb]
\centering
\caption{P-values and Cliff's deltas of the differences between different types of technical debt in terms of their removal rates among the introduced technical debt instances along the development process.
We report the pairs of SATD types with p-value $<$ 0.05 (i.e., significant) with $\star$ and present the interpretation of the corresponding Cliff's delta.
}\label{table_removal_rate_test}
\resizebox{\linewidth}{!}{
\begin{tabular}{c|cccccc}
       \toprule
       & \textbf{Algorithm} & \textbf{Compatibility} & \textbf{Defect}   & \textbf{Design}  & \textbf{Documentation} & \textbf{Requirement} \\
       \hline
Compatibility &                    &                        &                   &                  &                        &                      \\
Defect        &                    &                        &                   &                  &                        &                      \\
Design        & $\star$ (+medium)  & $\star$ (+large)       & $\star$ (+medium) &                  &                        &                      \\
Documentation &                    &                        &                   & $\star$          &                        &                      \\
Requirement   & $\star$ (+large)   & $\star$ (+large)       & $\star$ (+large)  &                  & $\star$ (+large)       &                      \\
Test          &                    &                        &                   & $\star$ (-small) &                        & $\star$ (-small)     \\
\bottomrule
\end{tabular}
}
\end{table}

Finally, to have an overview of the removal rate of the different types of technical debt in 7 deep learning frameworks before the latest stable release version, we calculate the proportion of removed technical debt instances among all the technical debt instances that are introduced over one year before the latest stable release version.
Table \ref{table_remove_ratio} shows the removal rate of different types of technical debt before the latest stable release version, as well as the average removal rate for each project.
To better view the differences between different types of technical debt, we highlight the removal rates of different types of technical debt which are higher than the corresponding project value in bold.

\begin{table}[!htb]
\centering
\caption{Removal rate of different types of technical debt}\label{table_remove_ratio}
\resizebox{\linewidth}{!}{
       \begin{tabular}{c|cccccccc}
       \toprule
       \textbf{Project   Name} & \textbf{TF}     & \textbf{Keras}   & \textbf{CNTK}   & \textbf{Caffe}  & \textbf{MXNet}  & \textbf{PyTorch} & \textbf{DL4J}    & \textbf{Average} \\
       \hline
       Project level           & 56.8\%          & 79.3\%           & 79.9\%          & 38.1\%          & 42.6\%          & 70.4\%           & 89.6\%           & 65.2\%           \\
       \hline
       Design                  & \textbf{68.5\%} & \textbf{99.0\%}  & \textbf{83.3\%} & \textbf{38.9\%} & \textbf{72.6\%} & \textbf{82.9\%}  & \textbf{93.8\%}  & \textbf{77.0\%}  \\
       Compatibility           & \textbf{60.1\%} & 56.5\%           & 74.8\%          & 29.6\%          & 0.0\%           & 67.4\%           &                  & 48.1\%           \\
       Defect                  & 51.3\%          & 50.0\%           & \textbf{79.6\%} & 27.3\%          & \textbf{61.5\%} & \textbf{77.1\%}  & 74.2\%           & 60.1\%           \\
       Documentation           & 50.0\%          &                  & \textbf{86.5\%} & \textbf{57.4\%} & 0.0\%           & \textbf{75.0\%}  & \textbf{100.0\%} & 61.5\%           \\
       Test                    & 50.0\%          & \textbf{100.0\%} & 78.5\%          & 28.6\%          & \textbf{60.0\%} & 55.6\%           & \textbf{93.3\%}  & \textbf{66.6\%}  \\
       Algorithm               & \textbf{57.6\%} & 70.4\%           & 75.2\%          & 30.0\%          & 32.0\%          & 59.6\%           & 85.7\%           & 58.6\%           \\
       Requirement             & \textbf{60.0\%} & \textbf{100.0\%} & \textbf{81.6\%} & \textbf{55.0\%} & \textbf{72.1\%} & \textbf{75.1\%}  & \textbf{90.6\%}  & \textbf{76.3\%}  \\
       \bottomrule
       \end{tabular}
}
\end{table}

\vspace{0.1cm}\noindent\textbf{Results:}
Table \ref{table_removal_rate_test} shows that the differences between requirement debt and all other types of technical debt except design debt are significant, and the effect sizes range from small (for test debt) to large (for other types of technical debt).
Figure \ref{figure_removal_rate_format} shows that the removal rates of requirement debt are one of the highest along the development process, e.g., in Keras, CNTK, Caffe, MXNet, and DL4J.
This shows that \textbf{requirement debt is the most removed technical debt along the development process}.
Table \ref{table_removal_prop_test} shows that the differences between requirement debt and other types of technical debt in terms of their proportion among the removed technical debt along the development process are significant, and the effect sizes range from medium to large.
Figure \ref{figure_removal_formation} shows that requirement debt is resolved in every development phase.
More specifically, along the development process, requirement debt has the second largest proportion of removed technical debt along the development process.
Developers put the resolution of requirement debt in a high priority.
In TensorFlow, Keras, CNTK, and PyTorch, the removal rate of requirement debt increase along with the development.
This indicates that developers resolve more requirement debt than the introduction.
However, in Caffe, MXNet, and DL4J, the removal rate of requirement debt decrease with fluctuation.
This shows that developers introduce more requirement debt than their resolution, and there is an accumulation of requirement debt in Caffe, MXNet, and DL4J.
We suggest the project managers should slow down the proposal of new requirements and wait for the resolution of requirement debt.
Finally, until the latest stable release version, the removal rates of requirement debt in different frameworks range from 55\% for Caffe to 100\% for Keras.

Table \ref{table_removal_rate_test} shows that the differences between design debt and all other types of technical debt except requirement debt are significant, and the effect sizes range from negligible (for documentation debt) to large (for compatibility debt).
Figure \ref{figure_removal_rate_format} shows that the removal rate of design debt is one of the highest along the development process across the studied deep learning frameworks.
This shows that \textbf{design debt is the second most removed technical debt along the development process}.
Table \ref{table_removal_prop_test} shows that the differences between design debt and other types of technical debt in terms of their proportion among the removed technical debt along the development process are significant and large.
Figure \ref{figure_removal_formation} shows that design debt has the largest proportion among the removed technical debt instances in most of the development phases.
This shows that developers paid their effort to the resolution of design debt the most along the development process.
For example, in Section \ref{introduced}, we observe that the design debt in Caffe is introduced the most at the beginning of the development process and is seldomly introduced after the beginning of the development process.
However, in the 5th and 6th development phases, design debt has the largest proportion among the removed technical debt.
Though developers seldomly admitted the sub-optimal trade-offs or decisions related to the design of code in the 5th and 6th development phases, they have to pay efforts to the resolution of design debt that is legacy in their past work.
In MXNet, though design debt has the largest proportion among the removed technical debt along the development process, the removal rates of design debt decrease with fluctuation after the 3rd development phase.
This shows that developers introduce more design debt than removal, and there is an accumulation of design debt.
We suggest the developers in MXNet pay more attention to the design of code.
Finally, until the latest stable release version, the removal rates of design debt range from 38.9\% in Caffe to 99.0\% in Keras.

Table \ref{table_removal_prop_test} shows that the differences between documentation debt and other types of technical debt in terms of their proportion among the removed technical debt along the development process are significant, and the effect sizes range from medium to large;
the differences between test debt and algorithm debt, design debt, requirement debt, and documentation debt are significant, and the effect sizes range from medium to large.
This shows that \textbf{Documentation debt has the smallest proportion among the removed technical debt instances in different development phases, followed by test debt}.
Along the development process, we can observe that documentation debt and test debt is removed in certain development phases in CNTK, MXNet, PyTorch, and DL4J.
One possible reason is that the number of introduced test debt and documentation debt is small, and limited attention paid by developers can result in a large proportion of test debt and documentation debt get removed.
Finally, until the latest stable release version, the removal rates of documentation debt in different frameworks range from 0\% for MXNet to 100\% for DL4J, and the removal rates of documentation debt in different frameworks range from 28.6\% for Caffe to 100\% for Keras.

\vspace{0.2cm}\noindent \fbox{\parbox[c]{1\linewidth} {\emph{Requirement debt is removed the most along the development process, followed by design debt.
Documentation debt has the smallest proportion among the removed technical debt instances in different development phases, followed by test debt.
}}}

\subsection{RQ3: Which types of technical debt are removed the fastest?}\label{remove_pace}

\vspace{0.1cm}\noindent\textbf{Motivation:}
In Section \ref{remove_rate}, we characterize the removal of different types of technical debt during the development process of different deep learning frameworks.
However, it is still unclear about how long does it take to be removed since the introduction of different types of technical debt.
In this section, we would like to characterize the removal of different types of technical debt along their lifecycle.

\vspace{0.1cm}\noindent\textbf{Approach:}
To characterize the removal of different types of technical debt, we perform a series of survival analyses.
Survival analysis can statistically analyze the expected duration time of the objects before an event, e.g., death in biological organisms and failure in mechanical systems \citep{miller2011survival}.
Survival analysis also can handle the case that an object does not have an event during the observation time (i.e., censored).
In this paper, survival analysis can characterize the removal of different types of SATD instances.
The SATD instances that are not removed before the latest stable release version are right-censored.

\begin{figure}
\centering
\includegraphics[width = \linewidth]{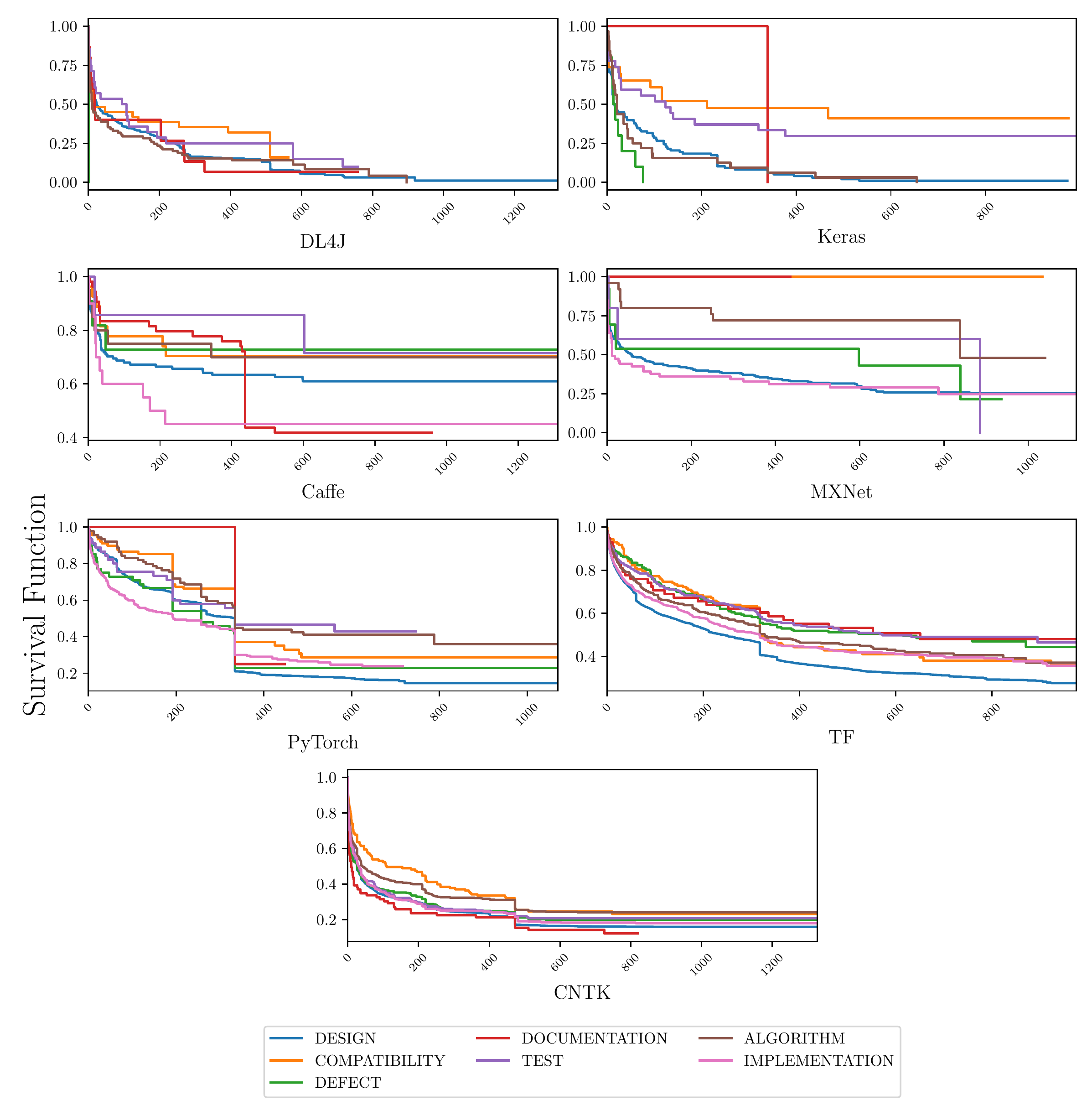}
\caption{Survival function of different types of SATD instances in all deep learning frameworks.
}\label{figure_survival} 
\end{figure}

To model the time to remove, survival function can give the probability that a SATD instance will survive beyond any specified time \citep{carpenter1997survival}.
We first estimate the survival function with popular parametric distributions, e.g., Exponential distribution, Weibull distribution, Gamma distribution, Log-Normal distribution, to find their best fit models.
Since the underlying data distribution is unknown, we use AIC to compare different models to select the most appropriate model for different types of technical debt in different deep learning frameworks.
The Akaike information criterion (AIC) can estimates the quality of each model, relative to each of the other models \citep{mcelreath2020statistical}.
As a result, we find that the survival time of certain types of technical debt in some deep learning frameworks cannot be significantly fit into any popular parametric distributions (i.e., p-values $>$ 0.05).
This motivates us to estimate the survival function with the Kaplan-Meier estimator \citep{kaplan1958nonparametric}.
The Kaplan-Meier estimator is a non-parametric statistic used to estimate the survival function from lifetime data.
Figure \ref{figure_survival} plots the survival function of different types of SATD instances in different deep learning frameworks.
Table \ref{table_median} presents the median survival time (i.e., half-life) of different types of technical debt in different frameworks.
To better view the differences between different types of technical debt, we highlight the median survival time of different types of technical debt which are shorter than the corresponding project value in bold.

\begin{table}
\centering
\caption{Half-life of different types of technical debt}\label{table_median}
\resizebox{\linewidth}{!}{
\begin{tabular}{c|ccccccc|c}
       \toprule
\textbf{Project   Name} & \textbf{TF} & \textbf{Keras} & \textbf{Caffe} & \textbf{PyTorch} & \textbf{MXNet} & \textbf{CNTK} & \textbf{DL4J} & \textbf{Average} \\
\hline
Project                 & 463.0       & 98.7           & 202.1          & 316.3            & 93.0           & 40.0          & 35.9          & 178.4            \\
\hline
Design                                   & \textbf{211.3}               & \textbf{25.7}                   & \textbf{81.3}                   & \textbf{270.0}                    & \textbf{65.2}                   & \textbf{25.6}                  & 31.3                           & \textbf{101.5}                             \\
Defect                                   & 604.2                        & \textbf{17.1}                   & \textbf{0.2}                    & \textbf{268.2}                    & INF                             & \textbf{21.8}                  & INF                            & 182.3                             \\
Compatibility                            & \textbf{420.9}               & INF                             &                                 & 317.6                             &                                 & 104.1                          &                                & 280.9                             \\
Requirement                              & \textbf{315.8}               & \textbf{21.3}                   & 477.0                           & \textbf{189.3}                    & \textbf{31.1}                   & \textbf{32.2}                  & \textbf{13.3}                  & \textbf{154.3}                             \\
Documentation                            & 660.7                        &                                 & 448.8                           & 334.1                             &                                 & \textbf{11.3}                  &                                & 363.7                             \\
Algorithm                                & \textbf{376.5}               & \textbf{89.8}                   & \textbf{3.1}                    & 432.3                             & 182.6                           & 50.9                           & 63.2                           & \textbf{171.2}                             \\
Test                                     & 651.4                        & 339.5                           & INF                             & 402.5                             & INF                             & \textbf{33.8}                  & 35.9                           & 292.6                             \\ 
\bottomrule
\end{tabular}
}
\end{table}

To describe the effect of different types of technical debt on the removal of SATD instances, we regress the types of technical debt against their hazard rate.
Hazard rate (i.e., failure rate) is used in survival analysis to describe the number of failures per unit of time.
We use Cox's model to estimate the hazard function (a function of time and some covariates that represent the hazard rate).
Cox's model is a non-parametric model to estimate the hazard function when the assumption of proportional hazards is true \citep{cox1984analysis}.
The proportional hazard assumption is that the shape of the hazard function is the same for all individuals and only a scalar multiple changes per individual.
We test the proportional hazard assumption using the scaled Schoenfeld residuals implemented in cox.zph() function in R.
We find there is no violation of the proportional hazard assumption in the survival time of different types of technical debt instances.
Table \ref{table_cox_test} presents the effects and their p-values of different types of technical debt on the removal of SATD instances.
The coefficients measure the impact (i.e., the effect size) of covariates.
Hazard ratios (HR) are calculated as the exponential of coefficients.
A hazard ratio above 1 indicates a covariate that is positively associated with the event probability, and thus negatively associated with the length of survival.

\begin{table}[]
\centering
\caption{Coefficients and hazard ratios of different types of SATD in Cox's model.
We report the coefficients with p-value $<$ 0.05 (i.e., significant) with $\star$.
}\label{table_cox_test}
\resizebox{\linewidth}{!}{
\begin{tabular}{c|ccccccc}
\toprule
& \textbf{design} & \textbf{requirement} & \textbf{defect} & \textbf{documentation} & \textbf{test} & \textbf{algorithm} & \textbf{compatibility} \\
\hline
\textbf{coef} & 0.44            & 0.23                 & 0.31            & 0.12                   & 0.01          & 0.08               & -0.08                  \\
\textbf{HR}   & 1.56            & 1.26                 & 1.36            & 1.13                   & 1.01          & 1.09               & 0.93                   \\
\textbf{p}    & $\star$         & $\star$              & $\star$         &                    &           & $\star$            &                 \\
\bottomrule
\end{tabular}
}
\end{table}

To check whether the differences in the survival time across the seven technical debt types are statistically significant, we perform a Wei-Lachin Test \citep{wei1984two}.
Wei-Lachin Test can compare survival distributions of more than two populations.
The null hypothesis that different types of technical debt instances have the same removal process.
As a result, we find that the difference between different types of technical debt in terms of survival time is significant (p-value $<$ 0.05).
Then, we perform a series of Mantel-Cox test as the post-hoc analysis to determine which groups differ from each other group \citep{mantel1966evaluation}.
Mantel-Cox Test can compare the survival distributions of two populations.
Table \ref{table_removalpace_test} presents p-values.

\begin{table}
\centering
\caption{P-values between different types of SATD in terms of the differences in survival time across the seven technical debt types.
We report the pairs of SATD types with p-value $<$ 0.05 (i.e., significant) with $\star$.
}\label{table_removalpace_test}
\resizebox{\linewidth}{!}{
\begin{tabular}{c|cccccc}
\toprule
              & Algorithm & Compatibility & Defect & Design & Documentation & Requirement \\ \hline
Compatibility & $\star$         &                     &              &        &               &                                  \\
Defect        & $\star$         & $\star$             &              &        &               &                                  \\
Design        & $\star$         & $\star$             & $\star$      &        &               &                                  \\
Documentation &                 & $\star$             &        & $\star$      &               &                                  \\
Requirement   & $\star$         & $\star$             &        & $\star$      &               &                                  \\
Test          &                 &                     & $\star$      & $\star$      &               & $\star$                                \\ 
\bottomrule
\end{tabular}
}
\end{table}

\vspace{0.1cm}\noindent\textbf{Results:}
Table \ref{table_removalpace_test} shows that the differences between design debt and all other types of technical debt are significant.
Table \ref{table_cox_test} shows a strong relationship between the design debt and shorter survival time, i.e., compared with comments that are not identified as SATD instances, design debt is removed 1.56 times faster.
This shows that \textbf{design debt is removed the fastest along the development process}.
Figure \ref{figure_survival} shows that along the lifecycle of design debt in different deep learning frameworks, design debt is one of the fastest removed technical debt.
In TensorFlow, design debt is removed the fastest along the lifecycle.
Table \ref{table_median} shows that the half-life of design debt range from 25.6 days (for CNTK) to 270.0 days (for PyTorch) with an average of 101.5.
Developers commonly put the resolution of design debt in the highest priority.

Table \ref{table_removalpace_test} shows that the differences between defect debt and algorithm debt, compatibility debt, design debt, and test debt are significant.
Table \ref{table_cox_test} shows a strong relationship between the defect debt and shorter survival time, i.e., compared with comments that are not identified as SATD instances, defect debt is removed 1.36 times faster.
This shows that \textbf{defect debt is the second-fastest removed technical debt along the development process}.
Figure \ref{figure_survival} shows that defect debt is one of the fastest removed technical debt.
In Keras and DL4J, defect debt is removed the fastest.
Table \ref{table_median} shows that the half-life of defect debt range from 0.2 days (for CNTK) to 270.0 days (for PyTorch) with an average of 101.5.
This shows that developers in Keras and DL4J put the resolution of defect debt in the highest priority.

Table \ref{table_removalpace_test} shows that the difference between compatibility debt and all other types of SATD except test debt is significant.
Table \ref{table_cox_test} shows there is no difference between compatibility debt and the comments that are not identified as SATD instances.
This shows that \textbf{compatibility debt is the slowest removed}.
We will discuss the compatibility debt in Section \ref{developer_how}.

Following that, Table \ref{table_removalpace_test} shows that test debt is significantly different from defect debt, design debt, and requirement debt;
documentation debt is significantly different from compatibility debt and design debt.
Table \ref{table_cox_test} shows there is no significant difference between documentation debt and the comments that are not identified as SATD instances;
there is no difference between test debt and the comments that are not identified as SATD instances.
This shows that \textbf{documentation debt and test debt is removed slow}.
The slow removal pace of documentation indicates that developers put the solving of documentation debt in low priority.
The slow removal pace of test debt can be associated with the un-accomplishment of the implementation of related functions to be tested.
We will discuss the slow removal of documentation debt and test debt in Section \ref{developer_how}.

\vspace{0.2cm}\noindent \fbox{\parbox[c]{1\linewidth} {\emph{Design debt is removed the fastest, followed by defect detb.
Compatibility debt is removed the slowest, followed by test debt and documentation debt.
}}}

\subsection{RQ4: Who removes different types of technical debt?}\label{who}

\vspace{0.1cm}\noindent\textbf{Motivation:}
Different from the technical debt that is hidden in code, SATD is a kind of technical debt that is acknowledged by the developers who write down the comments.
Previous work has illustrated that most of the SATD instances are self-removed, i.e., removed by the developers who introduce the comments \citep{maldonado_icsme2017}.
Moreover, technical debt also can be removed by other developers with more or fewer activities in the project.
More specifically, developers with more activities in the project are more likely to accomplish the more difficult programming tasks since they are more familiar with the project, while developers with fewer activities in the project are more likely to accomplish easier programming tasks since they contribute less to the project.
However, it is still unclear who removes the different types of technical debt the most during the development of deep learning frameworks.

\vspace{0.1cm}\noindent\textbf{Approach:}
To do so, we compare the author names and email addresses of the versions that introduce and remove the SATD instances to see if they are the same or not. 
\textbf{If the author's names and email addresses are the same in the version that introduces the SATD instance and the version that removes the SATD instance, the SATD instance is self-removed. Otherwise, the SATD instance is removed by other developers.}
Since there is the risk of misclassifying the authors that change their names in the source code repository during the evolution of the project, we rely on Open-hub's data to merge developer identities.
Hence, our study is only as accurate as Open-hub's classification.

If a SATD instance is removed by other developers, we compare the remover's activities in the project with the introducer's activities in the project at the removal time point.
We measure a developer's activities in the project by the number of commits performed by that developer in the given project.
More specifically, if the number of commits done by remover is more than the number of commits done by introducer at the removal of the SATD, then we consider the SATD to be removed by the developers with more activities in the project;
otherwise, the SATD is removed by developers with fewer activities in the project.

Table \ref{table_self_remove} presents the proportion of technical debt that is removed by developers with different activities in the project for different types of technical debt in different frameworks.
Concretely, we present the proportion of technical debt that is removed by other developers with more activities in the project (shortened as MORE), the proportion of technical debt that is removed by other developers with fewer activities in the project (shortened as FEWER), and the proportion of self-removed technical debt.

\begin{table}[!htb]
\centering
\caption{Proportion of Technical Debt Removed by Developers with Different activities in the project for Each Type}\label{table_self_remove}
\small
\resizebox{\linewidth}{!}{
\begin{tabular}{c|c|cccccccc}
\toprule
\textbf{Type}                   & \textbf{Removal Type} & \textbf{TF} & \textbf{Keras} & \textbf{CNTK} & \textbf{Caffe} & \textbf{MXNet} & \textbf{PyTorch} & \textbf{DL4J} & \textbf{Average} \\
\hline
\multirow{3}{*}{DESIGN}         & Self                  & 32.9\%      & 50.7\%         & 36.9\%        & 59.2\%         & 39.0\%         & 80.9\%           & 54.9\%        & 50.6\%           \\
                                   & More                  & 45.3\%      & 31.0\%         & 39.0\%        & 30.6\%         & 44.2\%         & 5.9\%            & 33.1\%        & 32.7\%           \\
                                   & Fewer                  & 21.8\%      & 18.3\%         & 24.1\%        & 10.2\%         & 16.9\%         & 13.2\%           & 12.0\%        & 16.6\%           \\ \hline
\multirow{3}{*}{COMPATIBILITY}  & Self                  & 17.6\%      & 0.0\%          & 21.0\%        & 50.0\%         &                & 87.7\%           &               & 35.3\%           \\
                                   & More                  & 69.2\%      & 100.0\%        & 58.0\%        & 37.5\%         &                & 8.8\%            &               & 54.7\%           \\
                                   & Fewer                  & 13.2\%      & 0.0\%          & 21.0\%        & 12.5\%         &                & 3.5\%            &               & 10.0\%           \\ \hline
\multirow{3}{*}{DEFECT}         & Self                  & 32.1\%      & 100.0\%        & 20.5\%        & 66.7\%         & 14.3\%         & 78.1\%           & 41.2\%        & 50.4\%           \\
                                   & More                  & 56.8\%      & 0.0\%          & 42.6\%        & 33.3\%         & 85.7\%         & 15.6\%           & 41.2\%        & 39.3\%           \\
                                   & Fewer                  & 11.1\%      & 0.0\%          & 36.9\%        & 0.0\%          & 0.0\%          & 6.3\%            & 17.6\%        & 10.3\%           \\ \hline
\multirow{3}{*}{DOCUMENTATION}  & Self                  & 38.9\%      &                & 22.6\%        & 19.4\%         &                & 100.0\%          &               & 45.2\%           \\
                                   & More                  & 44.4\%      &                & 38.7\%        & 67.7\%         &                & 0.0\%            &               & 37.7\%           \\
                                   & Fewer                  & 16.7\%      &                & 38.7\%        & 12.9\%         &                & 0.0\%            &               & 17.1\%           \\ \hline
\multirow{3}{*}{TEST}           & Self                  & 41.8\%      & 100.0\%        & 42.0\%        & 50.0\%         & 33.3\%         & 77.3\%           & 46.2\%        & 55.8\%           \\
                                   & More                  & 38.8\%      & 0.0\%          & 30.6\%        & 0.0\%          & 33.3\%         & 9.1\%            & 46.2\%        & 22.6\%           \\
                                   & Fewer                  & 19.4\%      & 0.0\%          & 27.4\%        & 50.0\%         & 33.3\%         & 13.6\%           & 7.7\%         & 21.6\%           \\ \hline
\multirow{3}{*}{ALGORITHM}      & Self                  & 33.3\%      & 42.1\%         & 29.8\%        & 66.7\%         & 28.6\%         & 70.0\%           & 52.4\%        & 46.1\%           \\
                                   & More                  & 49.1\%      & 47.4\%         & 43.0\%        & 33.3\%         & 71.4\%         & 10.0\%           & 33.3\%        & 41.1\%           \\ 
                                   & Fewer                  & 17.6\%      & 10.5\%         & 27.2\%        & 0.0\%          & 0.0\%          & 20.0\%           & 14.3\%        & 12.8\%           \\ \hline
\multirow{3}{*}{IMPLEMENTATION} & Self                  & 41.5\%      & 54.5\%         & 37.3\%        & 30.0\%         & 51.4\%         & 71.3\%           & 61.2\%        & 49.6\%           \\
                                   & More                  & 43.9\%      & 9.1\%          & 40.3\%        & 30.0\%         & 32.4\%         & 2.3\%            & 28.6\%        & 26.7\%           \\
                                   & Fewer                  & 14.6\%      & 36.4\%         & 22.4\%        & 40.0\%         & 16.2\%         & 26.4\%           & 10.2\%        & 23.7\%           \\ 
       \bottomrule

\end{tabular}
}
\normalsize
\end{table}

\begin{table}[]
\centering
\caption{P-values and Cliff's deltas of the differences between different developers who removed the SATD instances for each type of technical debt.
We report the pairs of SATD types with p-value $<$ 0.05 (i.e., significant) with $\star$ and present the interpretation of the corresponding Cliff's delta.
}\label{inner_test}
\resizebox{\linewidth}{!}{
\begin{tabular}{l|lllllll}
       \toprule
       & Algorithm & Compatibility   & Defect  & Design & Documentation & Requirement & Test \\ 
              \hline
Fewer - More & $\star$-small      & $\star$-medium        & $\star$-small   & $\star$-small   & $\star$-small          & $\star$-small        & $\star$       \\
Fewer - Self & $\star$-small      & $\star$-small         &                 & $\star$-small   &                        & $\star$-small        & $\star$-small \\
More - Self & $\star$            & $\star$+small         & $\star$+small   & $\star$         & $\star$+small          & $\star$              & $\star$-small \\
\bottomrule
\end{tabular}
}
\end{table}

To check the differences between different developers (i.e., developers with more activities in the project, developers with fewer activities in the project, and developers who introduce the SATD instances) in terms of their proportion among the removed SATD instances are significant for each type of technical debt, we perform seven Kruskal-Wallis H tests \citep{kruskal1952use}.
Kruskal-Wallis H test is a non-parametric test for comparing whether two or more independent samples originate from the same distribution.
As a result, we find that the difference between different developers in terms of their proportion among the removed SATD instances are significant (p-value $<$ 0.05).
Then, we perform a Dunn's test with Bonferroni correction to determine which groups differ from each other group (Dunn, 1961).
Dunn's test can be used for the post-hoc analysis for the specific sample pairs.
To calculate the effect size, we calculate the corresponding Cliff’s deltas \citep{cliff1993dominance}.
Cliff's delta is a measure of how often the values in one distribution are larger than the values in a second distribution.
Table \ref{inner_test} presents p-values and Cliff's deltas.

\begin{table}[]
\centering
\caption{P-values and Cliff's deltas of the differences between different types of technical debt in terms of whether the SATD instances are removed by developers with fewer activities in the project.
We report the pairs of SATD types with p-value $<$ 0.05 (i.e., significant) with $\star$ and present the interpretation of the corresponding Cliff's delta.
}\label{less_test}
\resizebox{\linewidth}{!}{
\begin{tabular}{c|cccccc}
\hline
              & Algorithm & Compatibility   & Defect  & Design & Documentation & Requirement \\ \hline
Compatibility & $\star$   &                 &         &        &               &             \\
Defect        & $\star$   & $\star$ (+small) &         &        &               &             \\
Design        &           & $\star$         & $\star$ &        &               &             \\
Documentation &           &                 &         &        &               &             \\
Requirement   &           &                 & $\star$ &        &               &             \\
Test          &           & $\star$         &         &        &               &             \\ \hline
\end{tabular}
}
\end{table}

\begin{table}[]
\centering
\caption{P-values and Cliff's deltas of the differences between different types of technical debt in terms of whether the SATD instances are removed by developers with more activities in the project.
We report the pairs of SATD types with p-value $<$ 0.05 (i.e., significant) with $\star$ and present the interpretation of the corresponding Cliff's delta.
}\label{more_test}
\resizebox{\linewidth}{!}{
\begin{tabular}{c|cccccc}
\hline
              & Algorithm & Compatibility   & Defect  & Design          & Documentation   & Requirement \\ \hline
Compatibility & $\star$   &                 &         &                 &                 &             \\
Defect        &           &                 &         &                 &                 &             \\
Design        & $\star$   & $\star$ (-small) & $\star$ &                 &                 &             \\
Documentation &           &                 &         & $\star$ (+small) &                 &             \\
Requirement   &           & $\star$ (-small) & $\star$ &                 & $\star$ (-small) &             \\
Test          & $\star$   & $\star$         & $\star$ &                 & $\star$ (-small) &             \\ \hline
\end{tabular}
}
\end{table}

\begin{table}[]
\centering
\caption{P-values and Cliff's deltas of the differences between different types of technical debt in terms of whether the SATD instances are removed by developers who introduce the SATD instances.
We report the pairs of SATD types with p-value $<$ 0.05 (i.e., significant) with $\star$ and present the interpretation of the corresponding Cliff's delta.
}\label{self_test}
\resizebox{\linewidth}{!}{
\begin{tabular}{c|cccccc}
\hline
              & Algorithm & Compatibility & Defect          & Design          & Documentation   & Requirement \\ \hline
Compatibility &           &               &                 &                 &                 &             \\
Defect        & $\star$   &               &                 &                 &                 &             \\
Design        & $\star$   &               & $\star$ (+small) &                 &                 &             \\
Documentation &           &               &                 & $\star$ (-small) &                 &             \\
Requirement   & $\star$   & $\star$       & $\star$ (+small) &                 & $\star$ (+small) &             \\
Test          & $\star$   & $\star$       & $\star$ (+small) &                 & $\star$ (+small) &             \\ \hline
\end{tabular}
}
\end{table}

To check the differences between different types of technical debt instances in terms their proportion that is removed by different developers (i.e., developers with more activities in the project, developers with fewer activities in the project, and developers who introduce the SATD instances) among the removed SATD instances are significant, we perform three Kruskal-Wallis H tests \citep{kruskal1952use}.
Kruskal-Wallis H test is a non-parametric test for comparing whether two or more independent samples originate from the same distribution.
As a result, we find that \textbf{the difference between different types of technical debt in terms of their proportion that is removed by different developers is significant (p-value $<$ 0.05)}.
Then, we perform a Dunn's test with Bonferroni correction to determine which groups differ from each other group \citep{dunn1961multiple}.
Dunn's test can be used for the post-hoc analysis for the specific sample pairs.
To calculate the effect size, we calculate the corresponding Cliff's deltas \citep{cliff1993dominance}.
Cliff's delta is a measure of how often the values in one distribution are larger than the values in a second distribution.
Table \ref{less_test}, Table \ref{more_test}, and Table \ref{self_test} present p-values and Cliff's deltas.

\vspace{0.1cm}\noindent\textbf{Results:}
Table \ref{inner_test} shows that the differences between the developers who introduce the SATD instances and other developers are significant and small in test debt.
Table \ref{self_test} shows that in terms of whether SATD instances are removed by developers who introduce them, documentation debt is significantly different from design debt, requirement debt, and test debt, and the effect sizes are small;
defect debt is significantly different from design debt, requirement debt, and test debt, and the effect sizes are small.
This shows that \textbf{documentation debt and defect debt is the least self-removed. Test debt, design debt, and requirement debt are the most self-removed}.
This shows that the developers who introduce the test debt, design debt, and requirement debt acknowledge the existence of the introduced technical debt.
They paid off these technical debt instances in their future work.
In contrast, the documentation debt and defect debt are removed the least by the developers who introduced them.
Table \ref{inner_test} shows that the differences between developers with more activities in the project and other developers are significant and small in defect debt and documentation debt, indicating that documentation debt and defect debt are removed more developers with more activities in the project.
One possible reason is that developers who introduce the defect debt and documentation debt may not know how to resolve these technical debt instances.

Table \ref{inner_test} shows that the differences between the developers with fewer activities in the project and others who removed the SATD instances are significant and small in algorithm debt, compatibility debt, design debt, and requirement debt.
Table \ref{less_test} shows that in terms of whether SATD instances are removed by developers with fewer activities in the project, the difference between compatibility debt and defect debt is significant and small.
This shows that \textbf{compatibility debt is removed the least by the developers with fewer activities in the project}.

Table \ref{inner_test} shows that the differences between the developers who introduce the SATD instances and the developers with more activities in the project are significant and small in compatibility debt, defect debt, documentation debt, and test debt.
Table \ref{more_test} shows that compatibility debt is significantly different from design debt, and requirement debt, and the effect sizes are small;
documentation debt is significantly different from design debt, test debt, and requirement debt, and the effect sizes are small.
This shows the \textbf{compatibility debt and documentation debt are the most removed by the developers with more activities in the project.}
One possible reason for compatibility debt is the resolution of compatibility debt may need the replacement of external dependencies.
However, the management of the update of external dependencies may require the privilege of administration in modern software management.
Besides, the implementation of the functions provided by external dependencies can require for the developers with more activities in the project.
Therefore, compatibility debt is removed the least by the developers with fewer activities in the project but is the most removed by the developers with more activities in the project.
One possible reason for documentation debt is that the documentation debt can be difficult for developers to resolve.
We will discuss the removal of documentation debt in Section \ref{developer_how}.

\vspace{0.2cm}\noindent \fbox{\parbox[c]{1\linewidth} {\emph{Documentation debt and defect debt is the least self-removed. Test debt, design debt, and requirement debt are the most self-removed.
Compatibility debt and documentation debt are the most removed by the developers with more activities in the project.
}}}

\section{Discussion}\label{discussion}

In this section, we depict the evolution of the frequencies of different types of technical debt along the development process, present our discussion on the removal patterns of different types of technical debt based on the findings we mentioned above, and provide actionable suggestions for practitioners, project managers, and researchers.
Finally, we present threats to validity.

\subsection{The Evolution of Frequencies of Different Types of Technical Debt along the Development Process}\label{evolution}

In Section \ref{introduced}, we investigate the introduction of different types of technical debt along the development process.
In Section \ref{remove_rate}, we investigate the removal of different types of technical debt along the development process.
However, it is still unclear the evolution of the frequencies of different types of technical debt along the developing process.

We plot the evolution of the frequencies of different types of technical debt along the development process.
To characterize the frequencies of different types of technical debt along the development process, we first divide the whole development process into ten \textbf{development phases} based on the chronological order of the commits.
Then we count the number of different types of technical debt instances in each development phase.
Figure \ref{figure_introduction_formation} shows the frequencies of different types of SATD instances along the developing process in 7 deep learning frameworks.
The frequencies of different types of technical debt at different development stages illustrate the challenges which are mainly confronted with by the developers at that time.
Ups and downs in the plots along the development process depict the changes in developers' challenges over time.

\begin{figure}[!htb]
\centering
\includegraphics[width = \linewidth]{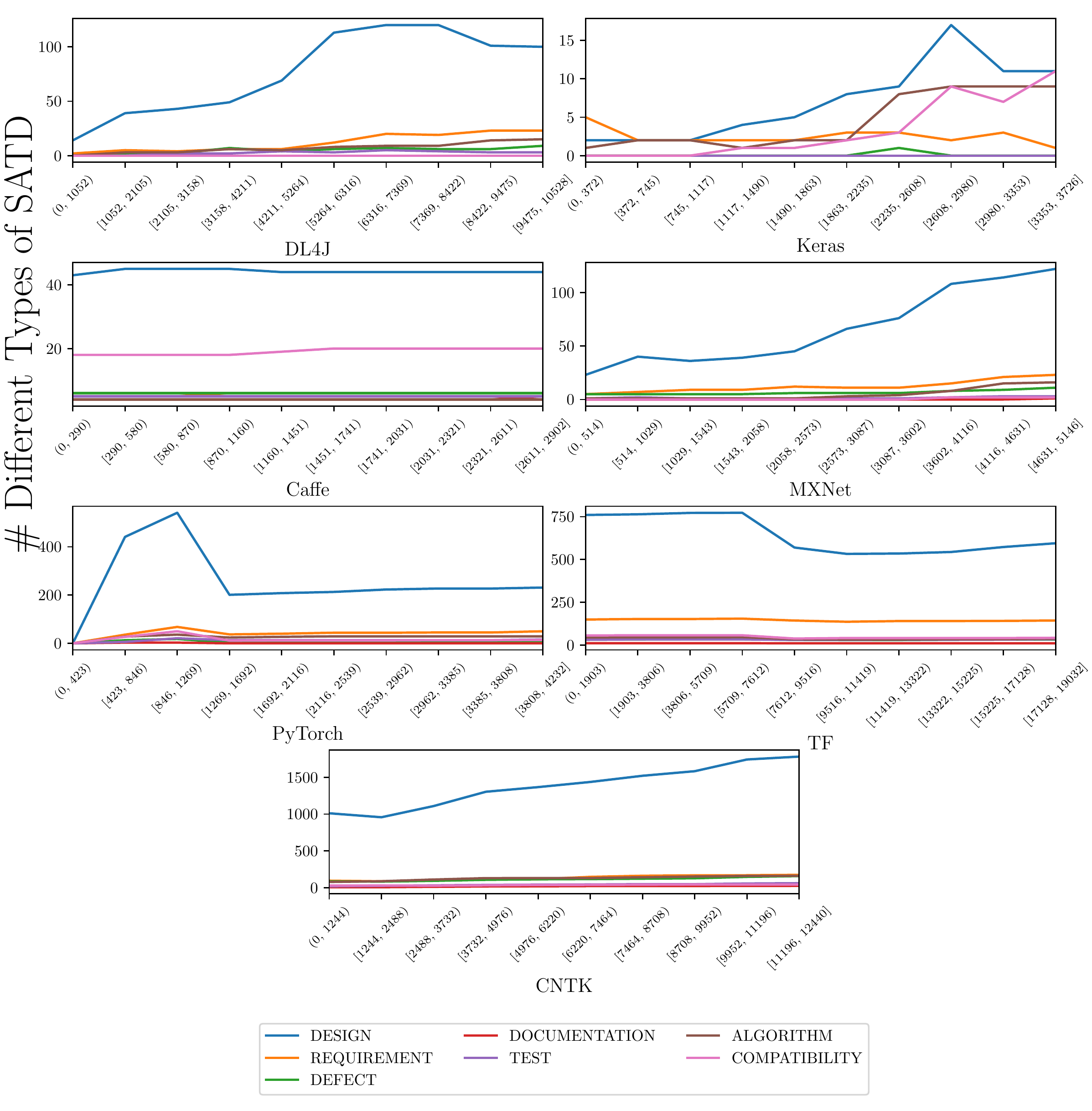}
\caption{Evolution of frequencies of different types of technical debt along the development process in different deep learning frameworks.}\label{formation} 
\end{figure}

Keras is an outlier here.
At the beginning of the development of Keras, requirement debt is the most common debt.
Developers are confronted with the fast iterate of the project and they record the unimplemented tasks.
Then the number of design debt instances increases along the development process, and design debt is the most common technical debt since the fourth development phase.
Many SATD instances are written down to express their dissatisfaction with the design of the implementation of tasks.
Meanwhile, the number of algorithm debt instances increases, and become the second most common technical debt.
It shows developers concentrate on the algorithms, i.e., the cutting edge deep learning module and the efficient computation method.
And currently, compatibility debt increases step by step and becomes one of the most common debt now.
This is because that Keras does not handle low-level operations such as tensor products, convolutions, and so on.
Instead, it relies on a specialized, well-optimized tensor manipulation library, e.g., Theano, to serve as the \emph{backend engine} of Keras. 
This leads to that Keras enjoys the convenience provided by backend projects at the cost of the maintenance of dependencies.

\subsection{Implications and Removal Patterns of Different Types of Technical Debt}\label{developer_how}

In this subsection, we discuss how different types of technical debt are removed based on the aforementioned findings.
Based on the discussion, we provide suggestions for developers, researchers, and project managers in the resolution of technical debt.

\noindent (1)  \textbf{Design debt:}
In Section \ref{introduced}, we observe that design debt is the most introduced technical debt along the development process.
In Section \ref{remove_rate}, we observe that design debt has the largest proportion among the removed technical debt instances in most of the development phases.
In Section \ref{remove_pace}, we observe that design debt is removed the fastest.

\noindent (2)  \textbf{Defect debt:}
In Section \ref{introduced}, Table \ref{table_introduced_test} shows that the proportion of defect debt is significantly lower than design debt and requirement debt.
One possible reason is the wide use of the industrial issue tracking system.
In Section \ref{who}, we observe that defect debt is one of the least self-removed.
In Section \ref{remove_pace}, we observe that the removal pace of defect debt is the second-fastest.
\textbf{We suggest that all developers participate in the resolution of defect debt.}

\noindent (3)  \textbf{Requirement debt:}
In Section \ref{introduced}, we observe that requirement debt is the second most common technical debt during the development process.
This shows that developers commonly cannot finish the implementation of certain tasks in time.
In Section \ref{who}, we observe that requirement debt is one of the most self-removed technical debt.
In Section \ref{remove_pace}, we observe that requirement debt is removed relatively fast.

\noindent (4)  \textbf{Documentation debt:}
In Section \ref{introduced}, we observe that the documentation debt is the least common technical debt.
This shows that developers seldomly admitted the sub-optimal trade-offs or decisions related to the documentation during the development process.
In Section \ref{who}, we observe that documentation debt is the least self-removed technical debt, and is the most removed by the developers with more activities in the project.
In Section \ref{remove_pace}, we observe that documentation debt is the second slowest removed technical debt.
This shows that compared with other developers, developers with more activities in the project can resolve documentation debt but not timely.
For example, the comment in TensorFlow:

\textit{``Given a numerical function ``f'', returns another numerical function ``g'', such that if ``f'' takes N inputs and produces M outputs, ``g'' takes N + M inputs and produces N outputs. I.e., if $(y_1, y_2, ..., y_M)$ = $f(x_1, x_2, ..., x_N)$, g is a function which is $(dL/dx_1, dL/dx_2, ..., dL-/dx_N)$ = $g(x_1, x_2, ..., x_N,dL/dy_1, dL/$-$dy_2, ..., dL/dy_M)$, where L is a scalar-value function of $(...x_i...)$. TODO(zhifengc): Asks math expert to say the comment again.''}

\noindent This comment shows that the documentation debt waits for the experts' resolution.
\textbf{We suggest that project managers allocate the documentation debt timely to the developers with more activities in the project.}

\noindent (5)  \textbf{Test debt:}
In Section \ref{introduced}, we observe that the test debt is one of the least common technical debt.
One possible reason is the widespread use of industrial test systems.
In Section \ref{who}, we observe that test debt is the most self-removed technical debt.
In Section \ref{remove_pace}, we observe that test debt is removed slower than other types of technical debt.
One reason is that the removal of test debts is associated with the accomplishment of the requirement to be tested.
For example, in a comment in PyTorch:

\textit{``TODO(jiayq): when there are backward and GPU implementations, enablethese two.self.assertDeviceChecks(dc, op, [X, scale, bias], [0])self.assertGra-dientChecks(gc, op, [X, scale, bias], 0, [0])''}

\noindent This shows that some requirement debt instances are not presented in a form of requirement debt but in the form of test debt.
\textbf{We suggest that project managers can check the unaccomplished requirement from the test debt.
}

\noindent (6)  \textbf{Compatibility debt:}
In Section \ref{who}, we observe that compatibility debt is removed the least by developers with fewer activities in the project, but is removed the most by developers with more activities in the project.
However, in Section \ref{remove_pace}, we observe that the removal of compatibility debt is the slowest.
One possible reason is that the removal of compatibility debts is associated with the release of qualified dependencies or the implementation of related functions.
For example, there is a comment in PyTorch:

\textit{``XXX: Gloo does not support scatter/gather/reduce''}

\noindent This comment indicates that the current code using Gloo\footnote{https://github.com/facebookincubator/gloo} to implement related tasks is sub-optimal.
However, this comment is not removed before the latest stable release version (i.e., May 31, 2018).
This is because that Gloo began to implement ``scatter/gather/reduce'' related code since Oct 2018\footnote{https://github.com/facebookincubator/gloo/commits/1d9e62aff9d7143129a69c8eb23e8351-e686ff3a/gloo/scatter.cc}, which is lagged behind the latest stable release version of PyTorch.
Dependencies that different projects rely on evolve at different paces, some of them remain unqualified until now.
\textbf{We suggest developers re-implement specific functions within the broader system architecture \citep{sculley2015hidden}.}
The re-implementation of the dependencies can make the frameworks not bind themselves tightly with its dependencies.

\noindent (7)  \textbf{Algorithm debt:}
In Section \ref{introduced}, we observe that the algorithm debt is the third most common technical debt.
This shows that for the development of deep learning frameworks, the sub-optimal trade-offs or decisions related to the algorithms are common.
In Section \ref{who}, Table \ref{inner_test} shows that algorithm debt is significantly less removed by developers with fewer activities in the project.
In Section \ref{remove_pace}, Table \ref{table_cox_test} and Table \ref{table_removalpace_test} shows that there is no significant difference between documentation debt and test debt, which is one of the slowest removed technical debt.
One possible reason is that the algorithms in deep learning frameworks are of a wide variety and still advancing, and newly proposed algorithms may be out of the range of developers' skill and difficult to implement. 
For example, in a comment in TensorFlow:

\textit{``Returns Poisson-distributed random number.  Uses Knuth's algorithm. Take care: this takes time proportional to lambda.  Faster algorithms exist but are more complex.''}

\noindent This shows that developers are aware of a faster algorithm to finish the development tasks, but they refuse to implement the algorithm due to complexity concerns.
\textbf{We suggest that all developers are involved in the fixing of algorithm debt.}

For \textbf{project managers}, Section \ref{evolution} depicts the evolution of the frequencies of different types of technical debt in different development phases along the development process.
These evolutions can reflect the changes of developers' concern at different stages along the development process.
For example, Section \ref{remove_rate} shows that though design debt has the largest proportion among the removed technical debt along the development process in MXNet, the removal rates of design debt decrease with fluctuation after the 3rd development phase.
This shows that developers introduce more design debt than removal, and there is an accumulation of design debt.
Project managers should find a balance between the quality of the project and the proposal of new requirements.
We suggest project managers take the evolution of the frequencies of SATD into consideration when managing their projects.

For \textbf{researchers}, we encourage future researchers to investigate the differences in the removal of the unresolved defects between the ones in the issue tracking system with the ones that are admitted as technical debt.
Our research finds that though issue tracking systems are used in the development process, there still is test debt and defect debt in source code.
However, it is unclear whether the unresolved defects attract less developer attention as compared to the ones in those issue tracking system systems.

We also encourage researchers to survey and categorize developers' intentions in resolving different types of technical debt.
In Section \ref{remove_pace}, our findings suggest that design debt is removed the fastest compared to all other types of technical debt, compatibility debt is removed the slowest.
However, the underlying reasons for developers to remove different types of technical debt in different paces are still uncleared.
We suggest further studies could survey and categorize developers' intentions when they resolving different types of technical debt.

\subsection{Threats to Validity}\label{threats}
Threats to internal validity concern factors that could have influenced our results. To identify SATD in a project, we use source code comments that describe part of the source code containing technical debt. One threat of using source code comments is the consistency of changes between the comments and the code, i.e., in some cases the comment may change but not the code and vice versa. However, previous work showed that between 72-91\% of the code and comment changes are consistent, i.e., code and comments co-change together \citep{Potdar_ICSME2014}.

To avoid bias caused by the SATD instances that are introduced recently before the latest stable version (i.e., right censoring) \citep{quesenberry1989survival}, we exclude the SATD instances that are introduced in one year before the latest stable release version.
One threat is that the SATD instances that are introduced most recently before the pruning-out window could still be subject to the right censoring problem and are not removed yet.
However, our findings show that 75\% of the SATD instances that are introduced before the latest stable version are removed in 299 days at the most (for PyTorch).
We believe only a small proportion of SATD instances that are introduced most recently before the pruning-out window are not removed.

To classify the detected source code comments into different types, we heavily depended on a manual process. Like any human activity, our manual classification is subject to personal bias and subjectivity. To reduce personal bias in manual classification of code comments, as we indicate in Section \ref{manually_classify}, the first author randomly sampled a statistically representative sample of 1,000 SATD instances from the 29,778 detected SATD instances using a 95\% confidence level with a 10\% confidence interval. We invite an independent Ph.D. student, who is not an author of this paper, to manually classify the randomly sampled 1,000 SATD instances.
The most common disagreement is that one technical debt instance can be associated with more than one category.
For example, in a comment in TensorFlow:

\textit{``This isn't strictly correct since in ghost batch norm, you are supposed to sequentially update the moving\_mean and moving\_variancewith each sub-batch. However, since the moving statistics are only used during evaluation, it is more efficient to just update in one step and should not make a significant difference in the result.''}

\noindent The first author labels this comment as an algorithm debt instance as this comment shows that the current implementation is a ``more efficient'' workaround.
In contrast, the independent Ph.D. student supposes this comment as a defect debt instance since this comment shows that the current implementation ``is not strictly correct'' in certain cases.
In fact, this SATD instance can be associated with two categories based on the interpretation of different people and can be labeled as defect debt and algorithm debt.
This shows that the labeling process is subject to personal bias and subjectivity.
For a certain technical debt instance, if we only focus on its introduction and removal as a certain category, there would be fewer SATD instances for other categories.
However, a high level of agreement between the classification given by the Ph.D. student and the first author is reported with Cohen's kappa coefficient of +0.79.
This gives us high confidence in the dataset used in our paper.

To identify the introduction and the removal of SATD instances, we consider the source code comments that do not exist anymore in a source code file as the removal of SATD.
The file where the comment exists being deleted also indicates that the comment does not exist.
However, in some cases, source code is partly moved from one file to another.
We treat this case as the removal of SATD in the original file and the introduction of SATD in the target file.

Moreover, we compare the author's names and email addresses to see if the SATD instances are self removed.
However, the authors can change their names in the source code repository during the evolution of the project.
To mitigate this threat, we rely on Open-hub's data to merge developer identities.
Hence, our study is only as accurate as Open-hub's classification.

Threats to external validity concern the generalization of our findings. Our study is conducted on seven large open source deep learning frameworks. Though we have discussed the similarities and differences between the deep learning frameworks and prior studies, our findings may not be generalized to other open source or commercial projects. In the future, we will analysis SATD in other systems.

We only discuss the test debt and the defect debt that are recorded in the source code.
However, issue tracking systems are used in the development process: our research does not consider the test debt and defect debt in issue tracking systems.
In the future, we plan to compare the test debt and defect debt reported in source code with that reported by means of issue tracking systems.

\section{Related Work}\label{related}
We divide our related work into two parts: the works on software engineering for deep learning and the research works on technical debt
We also compare the removal of technical debt in deep learning frameworks at the project-level with that in prior studies.

\subsection{Software Engineering for Deep Learning}

Considering the popularity and the importance of the deep learning projects, many researchers focus on developing solutions to help better engineer deep learning systems and libraries.

Many previous work focuses on the test of deep learning projects \citep{sun2018testing, sun2018concolic, ma2018deepgauge, zhang2018deeproad}.
For example, \citet{pei2017deepxplore} propose DeepXplore to systematically test DL systems and automatically identify erroneous behaviors without manual labels.
\citet{tian2018deeptest} propose DeepTest to automatically test DNN-driven autonomous cars, which can use test images that generated by different realistic transformations like rain, fog and lighting conditions.

Besides works focusing on testing of deep learning projects, Zhang et al.'s work \citep{zhang2018empirical} studies the characteristics of deep learning defects.
They study the TensorFlow application bugs from Stack Overflow and Github, and find the root causes of the defects, e.g., incorrect model parameter or structure.
\citet{islam19} investigate the bugs related to the five popular deep learning libraries, i.e., Caffe, Keras, TensorFlow, Theano, and Torch and find that data bug and logic bug are the most severe bug types.
Moreover, \citet{sculley2015hidden} empirically summarize the technical debt in machine learning systems during their development.
They explore several ML-specific risk factors in deep learning project design, including boundary erosion, entanglement, hidden feedback loops, undeclared consumers, data dependencies, and so on.

Different from those research works, our work inspects deep learning frameworks from another perspective: the removal of different types of technical debt.

\subsection{Technical Debt}

\citet{cunningham1993wycash} introduce the metaphor, technical debt, to describe the consequences of poor software development.
After the proposal of the metaphor, many researchers focus on how technical debt has been used to communicate the issues that developers find in the code in a way that managers can understand \citep{seaman2011measuring, kruchten2013technical, brown2010managing, lim2012balancing}.

Potdar et al.'s work \citep{Potdar_ICSME2014} uses comments to identify technical debt and nominates such technical debt as \textbf{self-admitted technical debt}.
They find that self-admitted technical debt is common cross projects and 2.4-31\% of the files in four traditional application projects contain such debt.
\citet{bavota2016large} replicate the study of Potdar et al.'s work \citep{Potdar_ICSME2014} on a large set of traditional application projects, i.e., Apache and Eclipse projects and find that approximately 57\% of self-admitted technical debts get removed and around 63\% of SATD are self-removed.
Maldonado et al.'s work \citep{maldonado_icsme2017} inspects the introduction and removal of self-admitted technical debt in five open source traditional application projects and find that the majority of self-admitted technical debt is removed (74.4\% on average) in 82 - 613.2 days on average.
\citeauthor{zampetti2018self} investigated whether SATD is ``accidentally'' removed, and the extent to which the SATD removal is being documented \citep{zampetti2018self}.
They observed that 8\% of the SATD removal is acknowledged in commit messages, and most of the changes addressing SATD require complex source code changes.

Many previous works also study the negative impact of technical debt during the development of projects.
Zazworka et al.'s work \citep{design_debt_impact} conducts a study to measure the impact of technical debt on software quality.
They find that god classes are more likely to change, and therefore, have a higher impact in software quality.
Fontana et al.'s work \citep{fontana2012investigating} investigates design technical debt and propose an approach to classify which code smell should be addressed first.
Wehaibi et al.'s work \citep{wehaibi2016examining} finds that self-admitted technical debt leads to more complex changes in the future development process. 

Many previous works also investigate the management of technical debt during the development of projects.
For example, \citeauthor{de2018aligning} proposed a framework for the prioritization of technical debt using a business-driven approach built on top of business processes \citep{de2018aligning}.
They also interviewed a set of IT business stakeholders, and collected and analyzed different sets of technical debt items, comparing how these items would be prioritized using a purely technical versus a business-oriented approach \citep{de2018aligning}.
\citeauthor{zampetti2020automatically} built a multi-level classifier capable of recommending six SATD removal strategies, e.g., changing API calls, conditionals, method signatures, exception handling, return statements, or telling that a more complex change is needed \citep{zampetti2020automatically}.

Moreover, the metaphor, technical debt, has been gradually extended to different types, e.g., design \citep{lim2012balancing}, and even documentation \citep{seaman2011measuring}, requirements \citep{ernst2012role}, and testing \citep{shull2011perfectionists}.
The work most relevant to us is Maldonado and Shihab's work \citep{Maldonado_MTD2015}, where they manually analyze the comments of 5 open source traditional application projects.
They find that there are five types of technical debt that are admitted in comments: design debt, defect debt, documentation debt, requirement debt and test debt.

Compared to these research works, our research focus on the removal of different types of technical debt in a family of software systems, i.e., the development of deep learning frameworks, where different frameworks are expected to achieve a same goal (i.e., offering high-level programming interfaces to deep learning applications) with the implementation of concrete tasks (e.g., implement core building blocks for designing, training and validating deep neural networks).
This enable us to find common patterns on the removal of technical debt.

\subsection{Comparison with Prior Studies}\label{compare}

Our research investigates technical debt in deep learning frameworks by analyzing the SATD in 7 open-source deep learning frameworks.
However, previous research \citep{maldonado_icsme2017} also investigates the removal of SATD on 5 open source traditional applications projects, i.e., Camel, Gerrit, Hadoop, Log4j, and Tomcat.
However, compared with our work, they perform an empirical study on the removal rate, survival time, and self-removal rate of the studies project \textbf{at the project level rather than the type level}.
Therefore, our study is not a replication study on another set of projects.
To compare the removal of SATD in deep learning frameworks with that in prior studies, we present their findings as well as the findings in our research in Table \ref{compare_table}.

\begin{table}[!htb]
       \centering
       \caption{Comparison between our findings and prior studies}\label{compare_table}
       \resizebox{\linewidth}{!}{
       \begin{tabular}{|p{0.2\linewidth}|p{0.33\linewidth}|p{0.38\linewidth}|}
       \hline
       \textbf{Topic}                     & \textbf{Prior Study}                                                                             & \textbf{Our Study}                                                                                     \\ \hline
       Proportion of removed SATD      & 74.4\% of SATD comments are removed on average. \citep{maldonado_icsme2017}.                                                                        & 67\% of the identified SATD comments is removed.                                                                                                                                          \\ \hline
       Survival time of removed SATD   & from 18.2 to 172.8 days on median. \citep{maldonado_icsme2017}.                                                     & from 9 to 95 days on median.                                                                                                                    \\ \hline
       Proportion of self-removed SATD & 54.4\% of SATD comments are self-removed \citep{maldonado_icsme2017}.                                                                              & the average self-remove rate is 42.19\%.                                                                                                                                                  \\ \hline
       \end{tabular}
       }
\end{table}

To check whether the differences between the removal of technical debt in deep learning frameworks and that in prior studies are statistically significant, we perform a Mann-Whitney U test \citep{mann1947test}.
As a result, the mean survival time of the technical debt in the deep learning frameworks is significantly shorter than that in prior studies (p-value $<$ 0.05).
This shows that\textbf{technical debt in deep learning frameworks is removed faster than the technical debt in traditional applications that is studied in prior work}.
Developers of deep learning frameworks are more active and put the solution of technical debt in higher priority compared with developers in traditional applications.
One possible reason is that a framework has a large user base, which puts greater pressure on the removal of technical debt.

Moreover, as indicated in Table \ref{compare_table}, compared with prior studies, SATD instances are removed relatively less in deep learning frameworks.
One possible interpretation is that developers of traditional applications put the resolution of technical debt in a relatively higher priority than the developers of deep learning frameworks.
Another possible interpretation is that technical debt in traditional applications are more easily resolved than deep learning frameworks.
In the future, we plan to compare the effort cost of the resolution technical debt in different types of projects (e.g., deep learning, traditional, IoT).
Furthermore, over half of the SATD is self-removed in traditional applications while less than half of the SATD is self-removed in deep learning frameworks.
This shows that the resolution of the technical debt in deep learning frameworks involve more developers

\section{Conclusion}\label{conclusion}
In this paper, we inspect the removal of different types of technical debt by mining SATD in the history version of 7 open source deep learning framework projects.
As a result, we find that developers admit design debt the most, and the removal rate of requirement debt is significantly higher than other types of technical debt.
Design debt is removed the fastest among all the types of technical debt, while compatibility debt is removed the slowest.
Documentation debt and defect debt is the least self-removed. Test debt, design debt, and requirement debt are the the most self-removed.
Compatibility debt and documentation debt are the most removed by the developers with more activities in the project.
Based on these findings, we depict the evolution of the frequencies of different types of technical debt along the development process.
In the future, we will examine the introduction and removal of technical debt with other evidence, such as by an interview.

\begin{acknowledgement}
TODO
\end{acknowledgement}

\balance
\bibliographystyle{spbasic}
\bibliography{myreference}

\end{document}